\documentclass[aps,pre,reprint,groupedaddress,showpacs,showkeys,floatfix]{revtex4-1}
\usepackage{graphicx, color}
\usepackage{amsmath}
\usepackage{amsfonts}
\usepackage{pgf}
\usepackage[breaklinks=true,colorlinks=true,linkcolor=blue,urlcolor=blue,citecolor=blue]{hyperref}
\usepackage[normalem]{ulem}
\newcommand{\qed}{\nobreak\ifvmode\relax\else
      \ifdim\lastskip<1.5em \hskip-\lastskip\hskip1.5em plus0em minus0.5em \fi \nobreak\vrule height0.75em width0.5em depth0.25em\fi}

\begin{document}

\title{Extreme events in Fitzhugh-Nagumo oscillators coupled with two time delays}

\author{Arindam Saha}
\email{arindam.saha@uni-oldenburg.de}

\author{Ulrike Feudel}
\email{ulrike.feudel@uni-oldenburg.de}

\affiliation{Theoretical Physics/Complex Systems, ICBM, University of Oldenburg, 26129 Oldenburg, Germany}



\begin{abstract}
We study two identical FitzHugh-Nagumo oscillators which are coupled with one or two different time delays. If only a single delay coupling is used, the length of the delay determines whether the synchronization manifold is transversally stable or unstable, exhibiting mixed mode or chaotic oscillations in which the small amplitude oscillations are always in-phase but the large amplitude oscillations are in-phase or out-of-phase respectively. For two delays we find an intricate dynamics which comprises an irregular alteration of small amplitude oscillations, in-phase and out-of-phase large amplitude oscillations, also called events. This transient chaotic dynamics is sandwiched between a bubbling transition and a blowout bifurcation.
\end{abstract}

\maketitle

\section{Introduction}

Extreme events have long been considered phenomena of crucial importance impacting various spheres of life~\cite{albeverio2006extreme}. Hence, extreme events have gained increasing attention in research areas ranging from social dynamics~\cite{RevModPhys.73.1067} oceanography~\cite{akhmediev2009waves, chabchoub2011rogue, chabchoub2012super, onorato2001freak, zakharov2006freak}, optics~\cite{bonatto2011deterministic, akhmediev2016roadmap, pisarchik2012multistate} and geophysics~\cite{sornette2003critical} to economics~\cite{bunde2002science, Feigenbaum2001} as well as power and communication grids~\cite{dobson2007}. Based on utility and relevance in the field under consideration, some studies focus on statistical properties of extreme events. In such cases, extreme events are characterized by `extreme value statistics'~\cite{gumbel2012statistics}. From a statistical perspective, extreme events occur in the tails of probability distributions that define the occurrence of events of a given size. Other studies, including the one presented here, approach extreme events from a dynamical systems perspective, where extreme events are defined as rare and recurrent events during which a specific dynamical variable exhibits an extremely large or small value~\cite{ansmann2013extreme, karnatak2014route}.

Previous investigations have revealed various mechanisms by which extreme events might be generated. For instance, finite dissipation in the complex Ginzburg-Landau equation leads to the generation of an incoherent background of interacting waves which occasionally initiates large amplitude events which can be considered as extreme events~\cite{kim2003statistics}. Localized excitations in arrays of coupled laser oscillators with random frequencies are known to wander in well defined trajectories, progressively increasing in synchrony as coupling strength is increased, and resulting in extreme events~\cite{rogister2007localized}. Optical rogue waves, another example of extreme events, are generated due to noise-induced attractor-hopping in multi-stable laser systems~\cite{pisarchik2011rogue, zamora2013rogue, reinoso2013extreme}. Studies performed with Integrate-and-Fire oscillators in a pulse-coupled small world network also show that chimera-states in such a network show irregular macroscopic dynamics with extreme synchrony under certain conditions~\cite{rothkegel2014irregular}.

Many physical systems that exhibit extreme events can be modeled as networks of excitable or oscillatory systems~\cite{ansmann2013extreme, kishore2011extreme}. Thereby the FitzHugh-Nagumo (FHN) model is widely used as a paradigmatic model which describes the dynamics of such systems on the nodes of the network. Investigations on networks of instantaneously coupled FHN oscillators with non-homogeneous parameters reveal that such networks are capable of generating extreme events~\cite{ansmann2013extreme, karnatak2014route} and a switching between various space-time patterns, extreme events being one of them~\cite{ansmann2016self}.

Here we analyze a system of two delay coupled FHN oscillators with identical parameters and show that delay-coupling can induce extreme events and hence, constitutes another generating mechanism of such events. The study has been motivated by various physical systems exhibiting extreme events where localized dynamical systems are connected via time-delayed connections. Examples include harmful algal blooms where concentrations of certain toxic plankton species occasionally increase to very high levels for a short interval of time. The resulting increase in concentrations of harmful, potentially toxic plankton species leads to a large impact on the ecosystem or even to human health. In addition to the local interplay between nutrients and competing species, oceanic currents which transport nutrients and species from one region of blooming activity to another are also important factors shaping the dynamics and transport of blooms in the ocean~\cite{bialonski2015data, bialonski2016phytoplankton}. In optical systems like lasers, rogue waves might be generated in systems where the finite travel time of signals may induce time delays~\cite{dal2013extreme}. Another example is neuronal communication between various regions of the brain which affect the levels of synchrony among these regions and lead to phenomena like epileptic seizures which are extreme events for the affected person~\cite{lehnertz2008epilepsy, Lehnertz2006}. In the latter case, communication between various regions of the brain is delayed due to the finite speed of signals through neurons. In general, complex systems with time delays are known to exhibit various spatio-temporal patterns~\cite{yanchuk2017spatio}. 

A common and important feature of some of the real-life networks with time-delayed coupling is the possibility of having more than one connection associated with the same pair of nodes in the network, each corresponding to a different time delay. For example, water currents originating from one region in the ocean might take two or more different paths to reach the same destination due to the presence of eddies or other barriers to hydrodynamic transport~\cite{bialonski2016phytoplankton}. Similarly, a neuronal signal originating from one region of the brain might take multiple pathways in the neuronal network to reach the same destination~\cite{atay2004stability, atay2006neural}. In this study, we focus on the consequences of having multiple delays in the coupling of components and its impact on the dynamics of the system. Thereby our main focus lies in the identification of the mechanism of generating extreme events.

The paper is organized as follows. In Sect.~2, we introduce the FHN model and the way how two FHN units are connected for this study. Furthermore, we discuss the emergence of additional fixed points due to coupling. Thereafter in Sect.~3, we analyze the qualitative dynamics of the system with respect to various parameter regimes. We start with investigating the long-term dynamics when only one delay is present and show the emergence of in-phase and out-of-phase large amplitude oscillation. Furthermore, we show the generation of extreme events upon the introduction of a second delay and classify the observed extreme events as belonging to two categories based on phase synchrony. We further demonstrate that the emergence of extreme events is not restricted to a single parameter set but occurs in a whole strip in the parameter plane spanned by the second delay time and the corresponding coupling strength. This strip is bounded by a bubbling transition on one side and a blowout bifurcation on the other side. We finally discuss the implications of the results obtained in Sect.~4 and present an outlook for future research.

\section{The Model and The Structure of The Phase Space}

\begin{figure}
\includegraphics[width=0.5\linewidth]{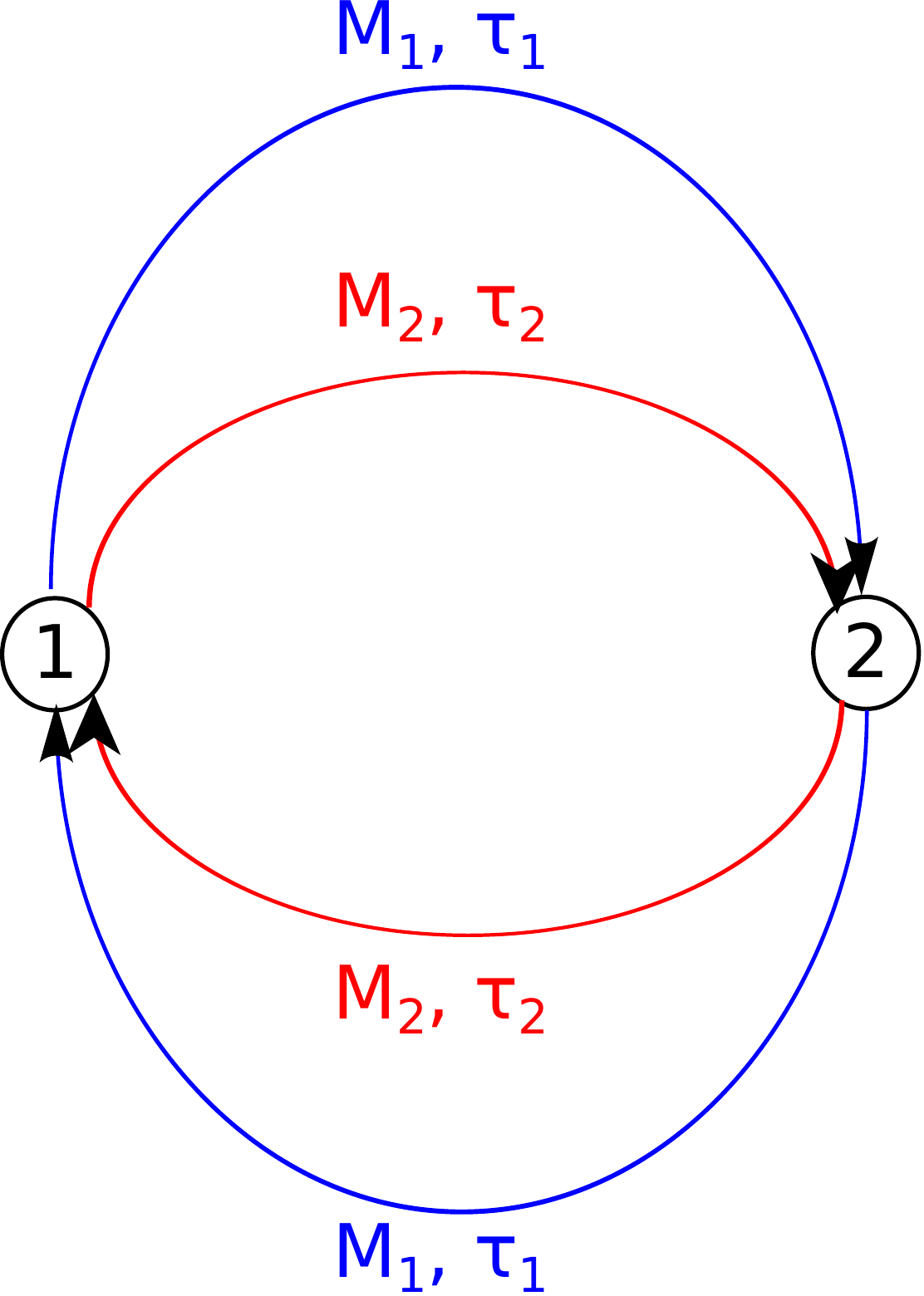}
\caption{(Color Online) Schematic representation of couplings between the two FHN units.}
\label{fig:Network}
\end{figure}

Let us consider a system of two coupled FHN units (Fig.~\ref{fig:Network}) whose dynamics is given as
\begin{equation}
\begin{aligned}
\dot{x}_i &= x_i(a_i-x_i)(x_i-1)-y_i + \sum_{k=1}^L M_k (x_j^{(\tau_k)}-x_i) \\
\dot{y}_i &= b_i x_i - c_i y_i + \sum_{k=1}^L M_k (y_j^{(\tau_k)}-y_i). 
\end{aligned}
\label{eq:Model}
\end{equation}
where $i,j \in \{1,2\}$, $i \neq j$ and,
\begin{equation}
\begin{aligned}
x_i^{(\tau_k)} &= x_i(t-\tau_k) \\
y_i^{(\tau_k)} &= y_i(t-\tau_k).
\end{aligned}
\end{equation}
Here the two FHN units -- with internal parameters $a_i$, $b_i$ and $c_i$ --- are coupled to each other by $L$ delay-couplings. Each delay coupling is characterized by its delay time $\tau_k$ and coupling strength $M_k$. For our study, we chose the internal parameters of the independent units to be identical; i.e. $a_1=a_2=a=-0.025$, $b_1=b_2=b=0.00652$ and $c_1=c_2=c=0.02$.  These values are chosen such that in absence of any coupling, the FHN units are in the oscillatory regime. Furthermore, for the numerical results presented, we assume that for all times $t<0$, the units have same values as the initial condition. The results for some other functional forms of histories were also checked and were found to give identical results in the long term limit.

Having identical parameters for individual FHN oscillators implies that the phase space is partitioned into two symmetric halves by an invariant synchronization manifold defined by $x_1(t-\tau)=x_2(t-\tau);~y_1(t-\tau)=y_2(t-\tau)$ for $0 \le \tau \le \text{max} \{ \tau_k \}$. In particular, if the system possesses a fixed point $(x_1,y_1,x_2,y_2)$ which lies not on the synchronization manifold, then the point $(x_2,y_2,x_1,y_1)$ is another fixed point with the same stability. Moreover, the symmetry of the system forces any fixed point away from the synchronization manifold to occur in pairs. Note that in absence of any coupling, the only fixed point of the system is the origin located on the invariant manifold. We now show that additional fixed points might be created in the system for sufficiently strong couplings.

For any fixed point $(x_1^*,y_1^*,x_2^*,y_2^*)$, we must have $\dot{x}_i^*=\dot{y}_i^*=0$ and $x_i^{*(\tau_k)}=x_i^*$, $y_i^{*(\tau_k)}=y_i^*$ for all $i$ and $k$. Imposing these conditions on Eq.~\eqref{eq:Model} and setting $L=2$ we get,
\begin{equation}
\begin{aligned}
x_1^*(a-x_1^*)(x_1^*-1)-y_1^* + (M_1+M_2) (x_2^*-x_1^*) &= 0 \\
b x_1^* - c y_1^* + (M_1+M_2) (y_2^*-y_1^*) &= 0 \\
x_2^*(a-x_2^*)(x_2^*-1)-y_2^* + (M_1+M_2) (x_1^*-x_2^*) &= 0 \\
b x_2^* - c y_2^* + (M_1+M_2) (y_1^*-y_2^*) &= 0.
\end{aligned}
\label{eq:Fixed_Point}
\end{equation}

\begin{figure}[t]
\includegraphics[width=\linewidth]{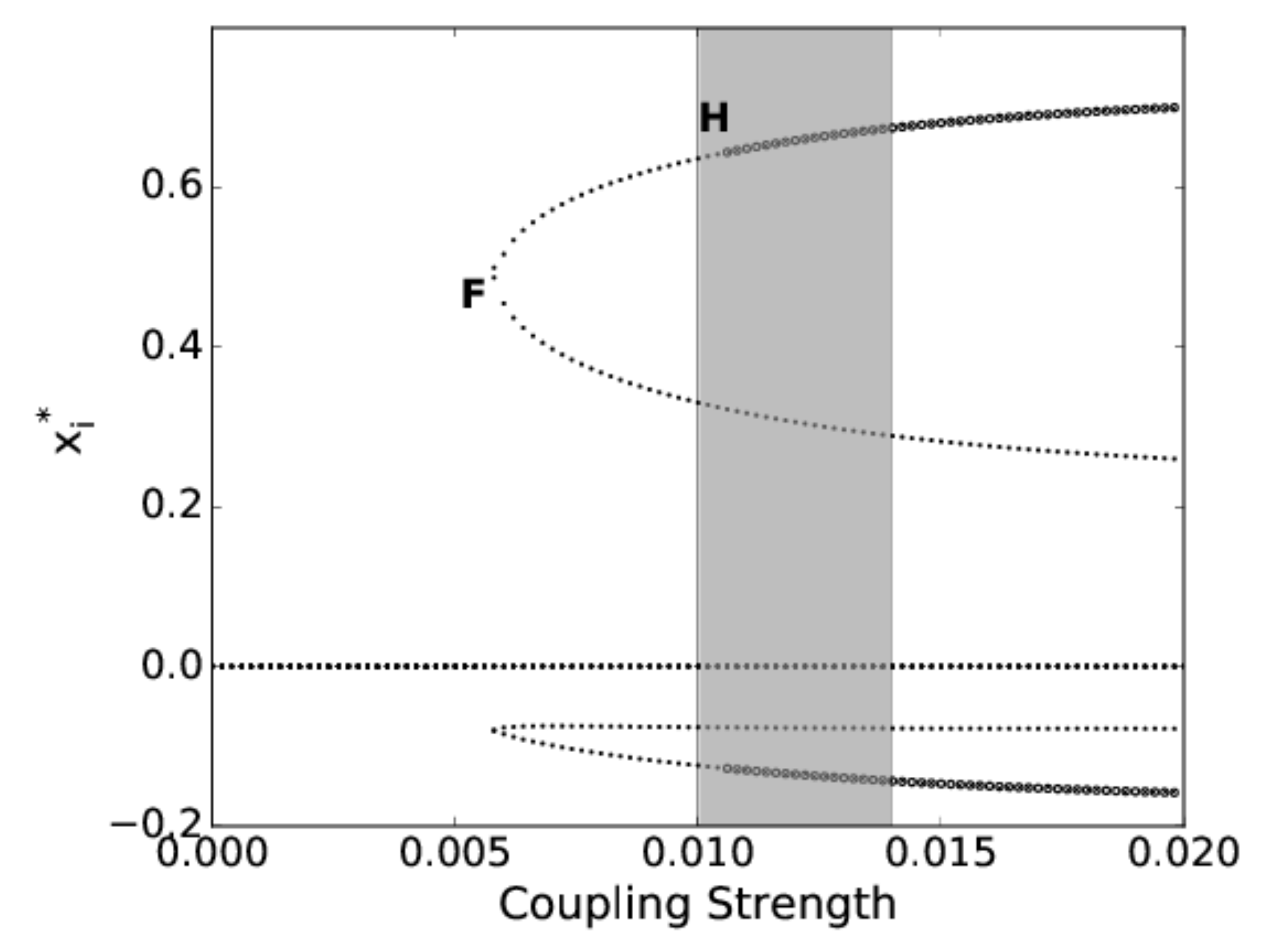}
\caption{Bifurcation diagram showing the position of the fixed points for varying coupling strength $M$. The filled circles represent stable fixed points and the small dots represent their unstable counterparts. In addition to the unstable origin the plot shows the creation of two pairs of unstable fixed points via a fold bifurcation \textbf{F} at $M \approx 0.0058$. One of these pairs stabilizes due to a Hopf bifurcation \textbf{H} at $M \approx 0.0105$ (denoted by \textbf{H}). The region in gray denotes the default parameter range chosen for our analysis. For this plot the FHN units using are connected by a single delay with $\tau=80$.}
\label{fig:FixedPoints}
\end{figure}

Note that the position of the fixed point is independent of the delays $\tau_k$ and depends on the total coupling strength, $M=M_1+M_2$, rather than the individual ones. For $M$ values close to zero, we have the origin as our only fixed point. However, as $M$ exceeds a threshold (see Fig.~\ref{fig:FixedPoints}), two unstable pairs of fixed points emerge simultaneously via a fold bifurcation. If $M$ is increased further, one of the pairs gets stabilized by a reverse Hopf bifurcation.

In order to investigate the dynamics of the system both in absence and presence of the non-trivial stable fixed points, we vary $M$ from 0.01 to 0.014 in this study.

\section{Dynamical Features of the System}

In this section we describe the general qualitative dynamics exhibited by the coupled FHN units for different coupling strengths and delay times. Since the range of possible dynamics that can be observed in this system is very diverse, we focus on the dynamical features that help us understand the emergence of extreme events in such systems. Before studying systems with two delays where extreme events appear, we first analyze single-delay systems. The dynamics of the latter are similar to two-delay systems but easier to understand.

Throughout this section, we use the terms `large amplitude oscillations' and `small amplitude oscillations' in the following sense: `Large amplitude oscillations' refer to oscillations with amplitude of the $x$-component larger than 0.5, otherwise they are called `small amplitude oscillations'.

\subsection{Single Delay}

We now study the dynamics of FHN units which are coupled by a single time delay. Setting the number of couplings, $L=1$ and dropping the subscript $k$ in Eq.~\eqref{eq:Model} we have,
\begin{equation}
\begin{aligned}
\dot{x}_i &= x_i(a-x_i)(x_i-1)-y_i + M (x_j^{(\tau)}-x_i) \\
\dot{y}_i &= b x_i - c y_i + M (y_j^{(\tau)}-y_i)
\end{aligned}
\end{equation}
with $i,j \in \left\{ 1,2 \right\}$.

For such a system, the synchronization manifold is either transversally stable or unstable as discussed in the following:

\subsubsection{Transversally Stable Synchronization Manifold}

\begin{figure}
\includegraphics[width=0.9\linewidth]{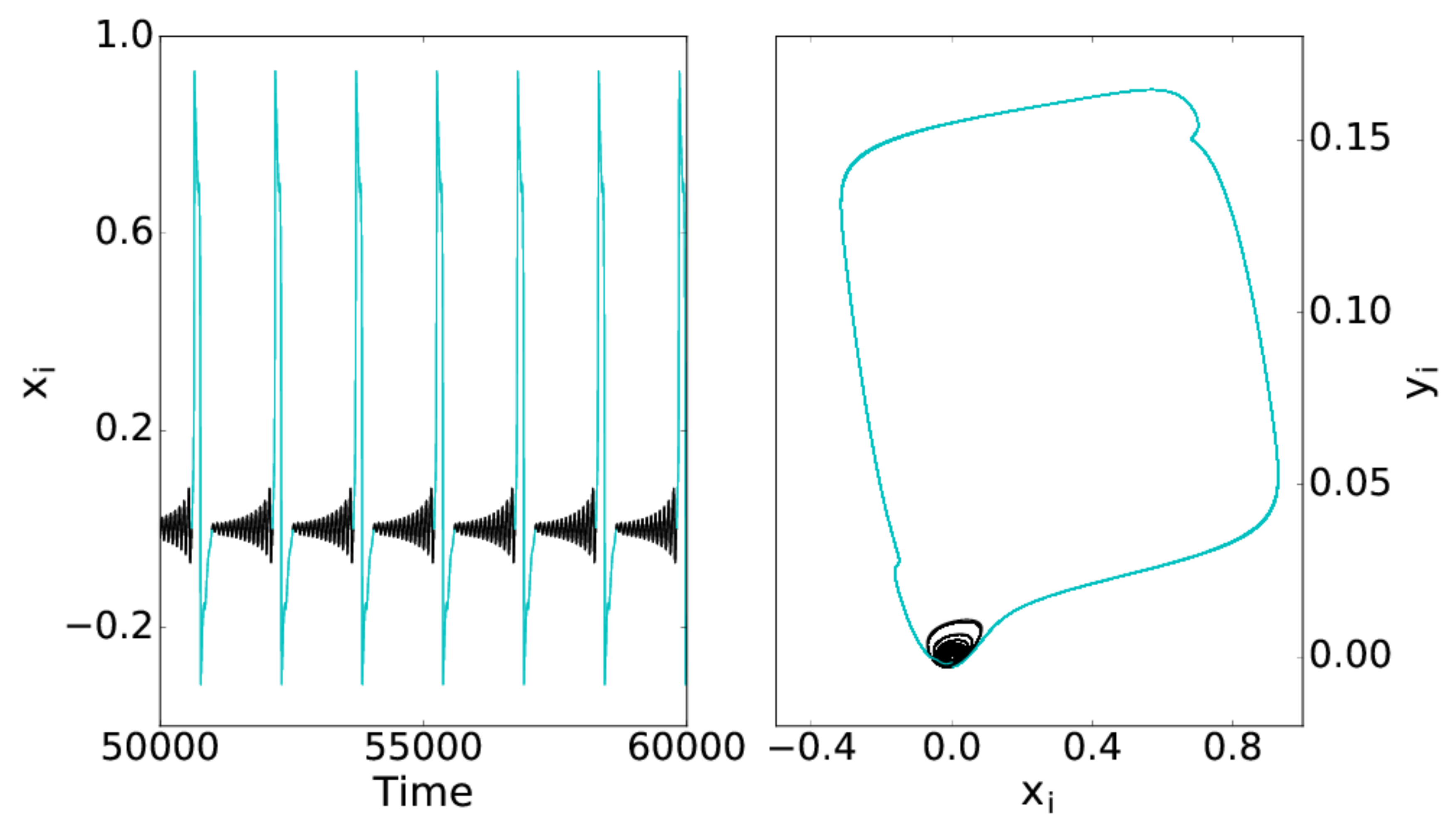}
\includegraphics[width=0.9\linewidth]{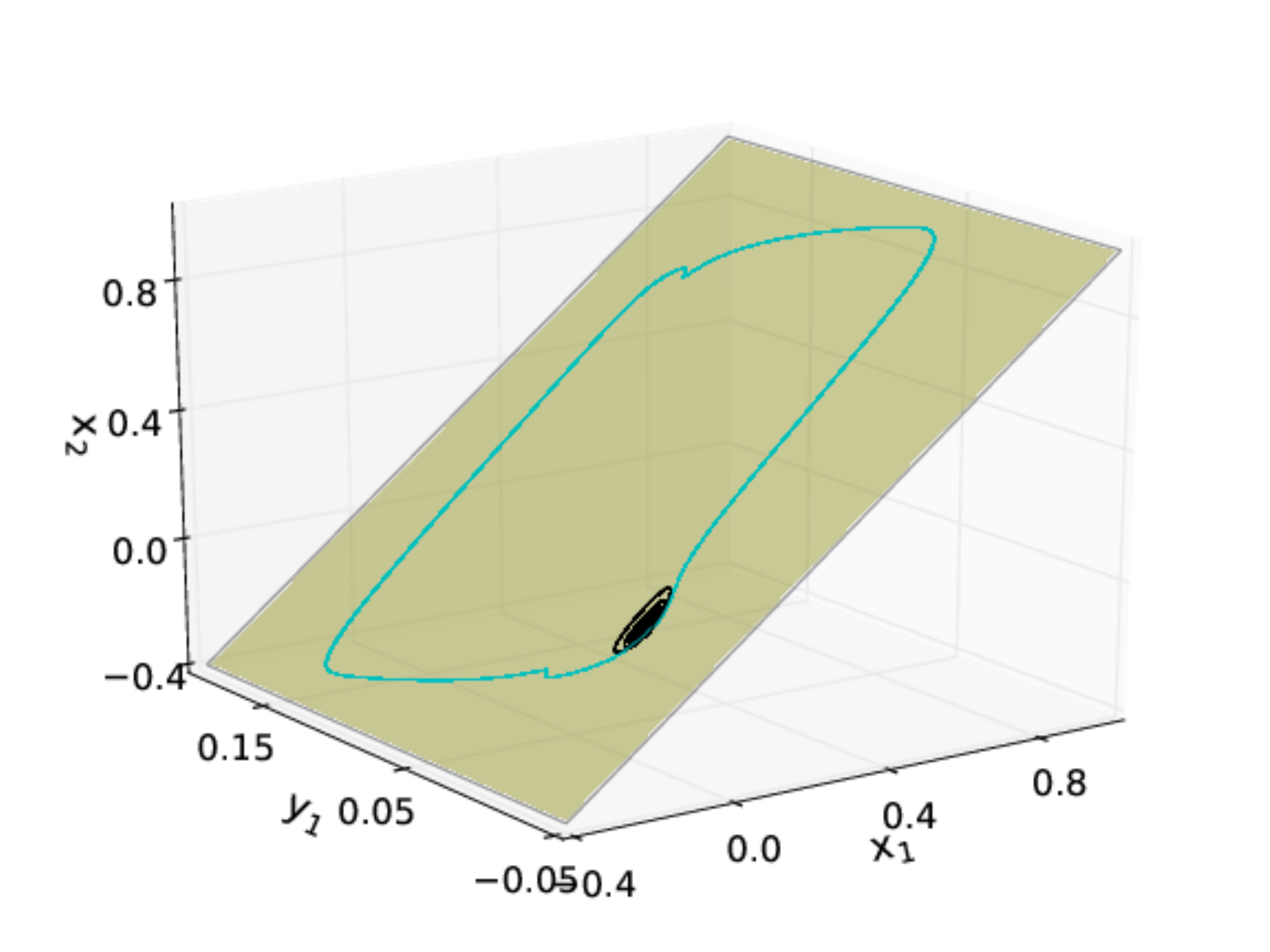}
\caption{(Color Online) Various representations of the synchronized long-term dynamics of two FHN units coupled with a single delay when the synchronization manifold is stable: top-left panel: time-evolution of the $x$-coordinates of the oscillators; top-right panel: trajectories of the two oscillators in phase space; bottom panel: trajectory in a 3-dimensional representation. The plane is the synchronization manifold. Color code: Small amplitude oscillations in black; synchronous large amplitude oscillations in cyan (light gray). Parameters: $M=0.01$, $\tau=80$.
}
\label{fig:SingleLargeDelay}
\end{figure}

\begin{figure}
\includegraphics[width=0.9\linewidth]{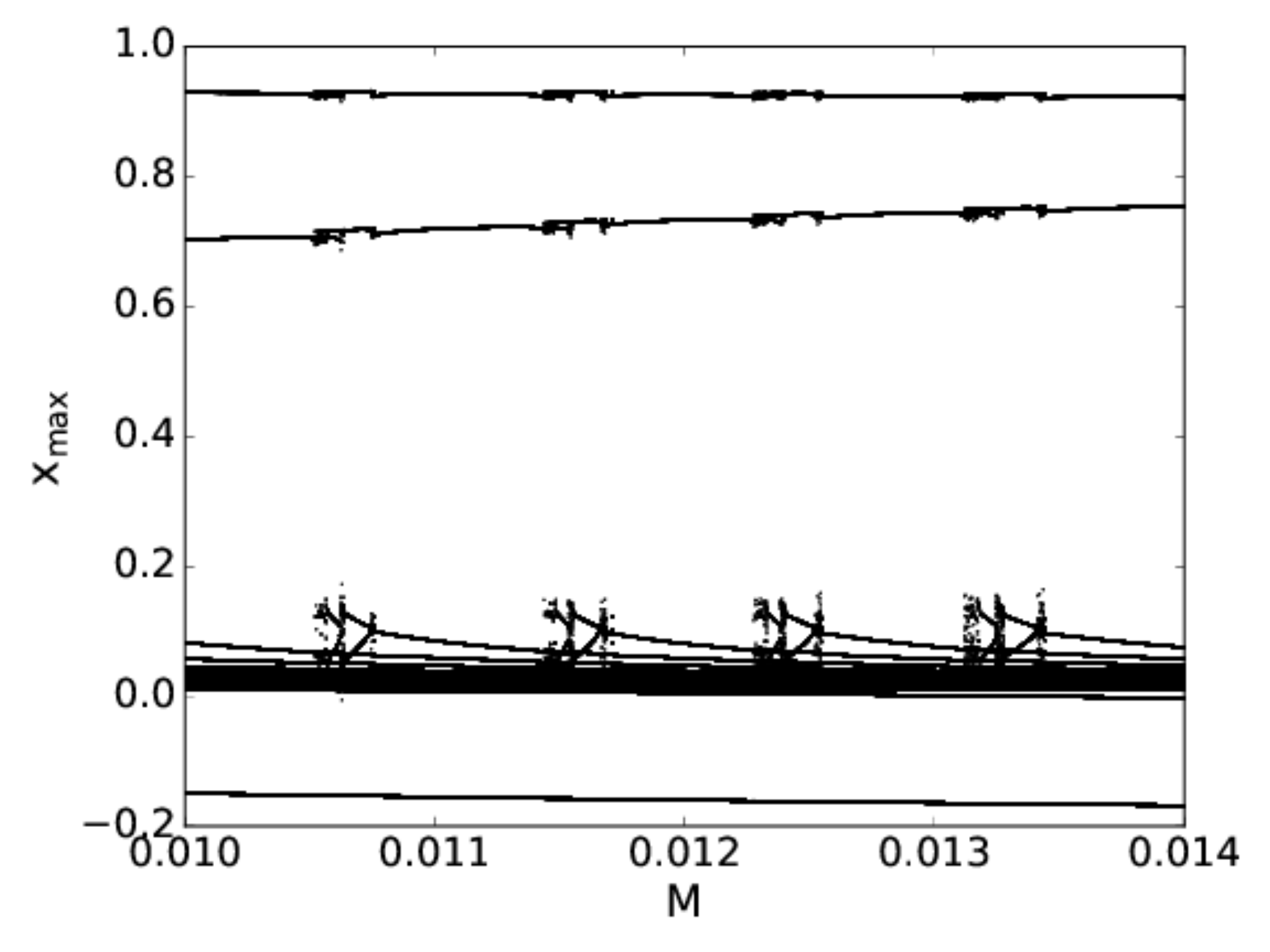}
\includegraphics[width=0.9\linewidth]{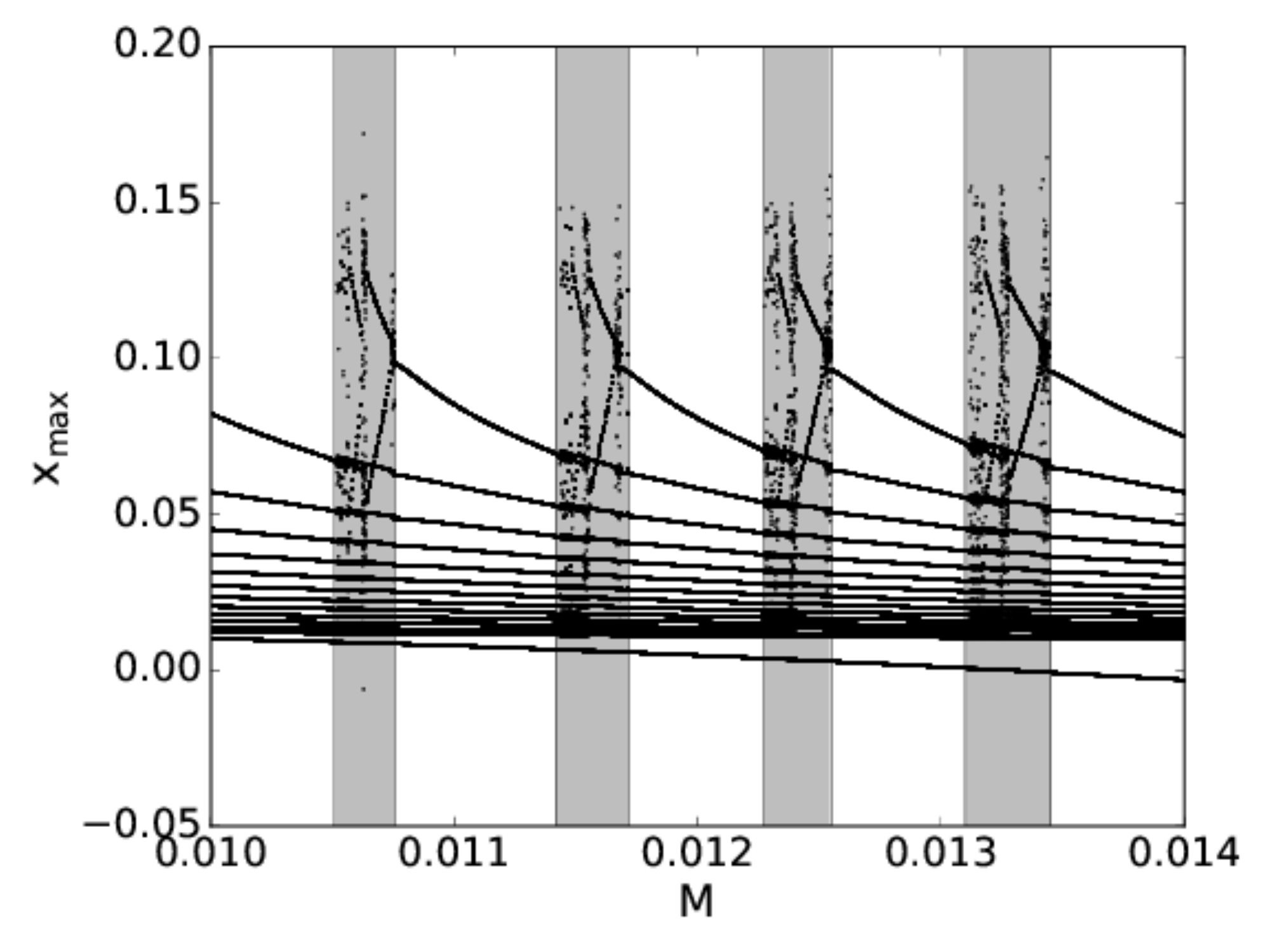}
\caption{Bifurcation diagram illustrating the period adding cascade observed with increasing coupling strength $M$ and fixed time-delay $\tau=80$. On the vertical axis the maximas corresponding to the oscillations are plotted. The upper panel shows the complete diagram, while the lower panel shows close-up view of the regions corresponding to small amplitude oscillations. The gray regions in the lower panel correspond to the chaotic dynamics which intersperse the mixed mode oscillations.}
\label{fig:PeriodAdding}
\end{figure}

For $\tau=80$, $M=0.01$, there are no stable fixed points in the system, the synchronization manifold is stable and contains the only attractor of the system. In such a case, the oscillators converge to the invariant synchronization manifold and execute mixed-mode oscillations on it as shown in Fig.~\ref{fig:SingleLargeDelay}. The oscillators are in complete synchrony and spend large time intervals near the origin executing small amplitude oscillations (shown in green) around it. During these oscillations, the units spiral away from the origin and their amplitude of oscillation increases over time. After a fixed number of small amplitude oscillations, the units execute one large amplitude oscillation (shown in blue). The amplitude of such an oscillation is more than five times larger than the largest of the small amplitude oscillations. At the end of this oscillation, the system ends up close to the origin and the small amplitude oscillations begin again. Note that the entire long term dynamics --- comprising of small and large amplitude oscillations --- occurs on the invariant synchronization manifold.

As the coupling strength $M$ is increased, the system undergoes a sequence of period-adding bifurcations (see Fig.~\ref{fig:PeriodAdding}). At each such bifurcation, the number of small amplitude oscillations separating any two large amplitude oscillations increases by one. The sequence of period adding is interspersed with intervals of chaotic dynamics, i.e.\ the two oscillators exhibit completely synchronized chaotic dynamics on the synchronization manifold.

When the coupling strength $M$ is increased even further such that a pair of fixed points becomes stable, the system shows multistability with some of the trajectories converging to one of the fixed points. However any trajectory that comes sufficiently close to the invariant synchronization manifold will never converge to the fixed points as it is attracted to that manifold and forever executes mixed mode or chaotic oscillations as described above.

\subsubsection{Transversally Unstable Synchronization Manifold}

\begin{figure}
\includegraphics[width=0.9\linewidth]{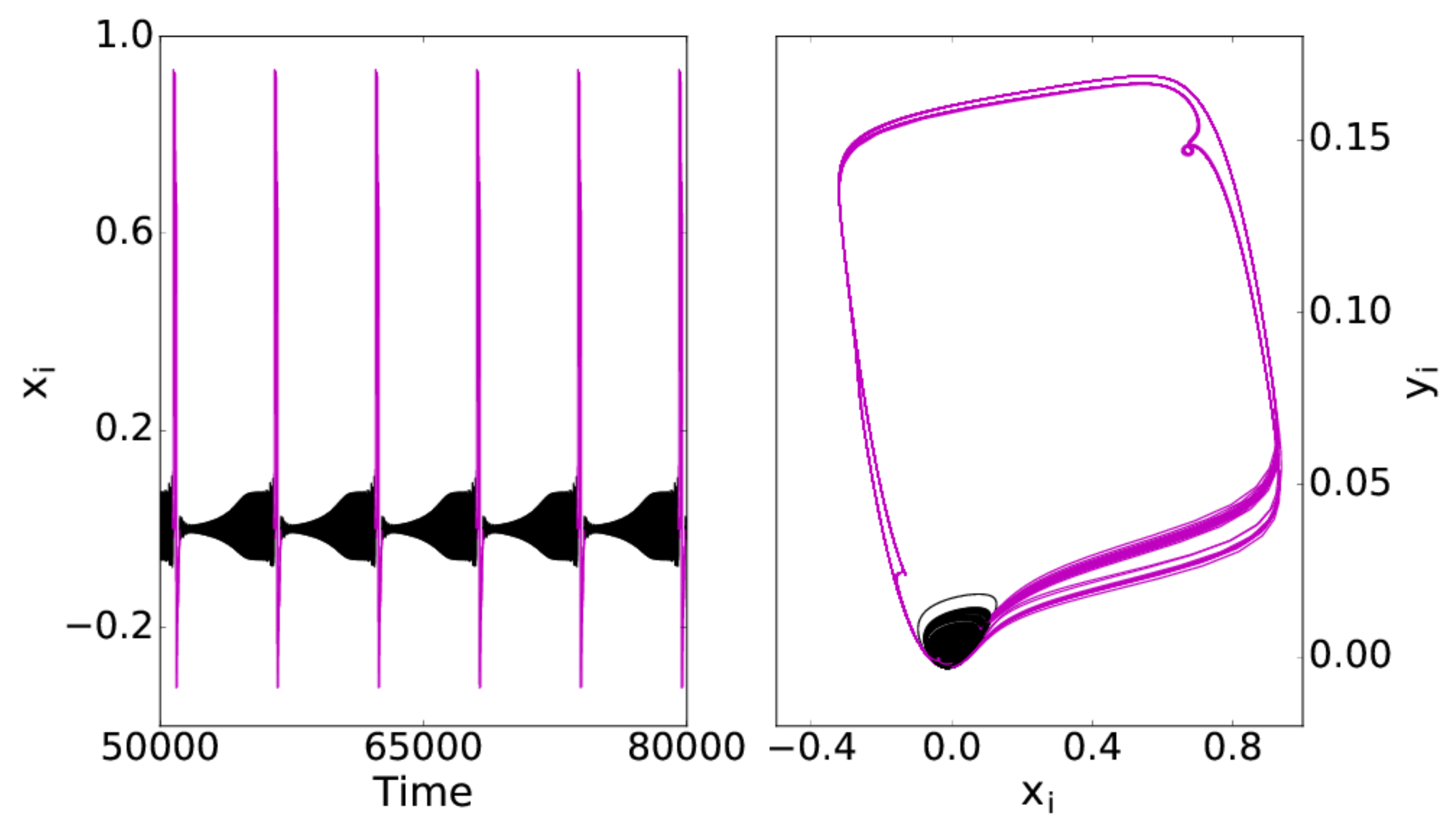}
\includegraphics[width=0.9\linewidth]{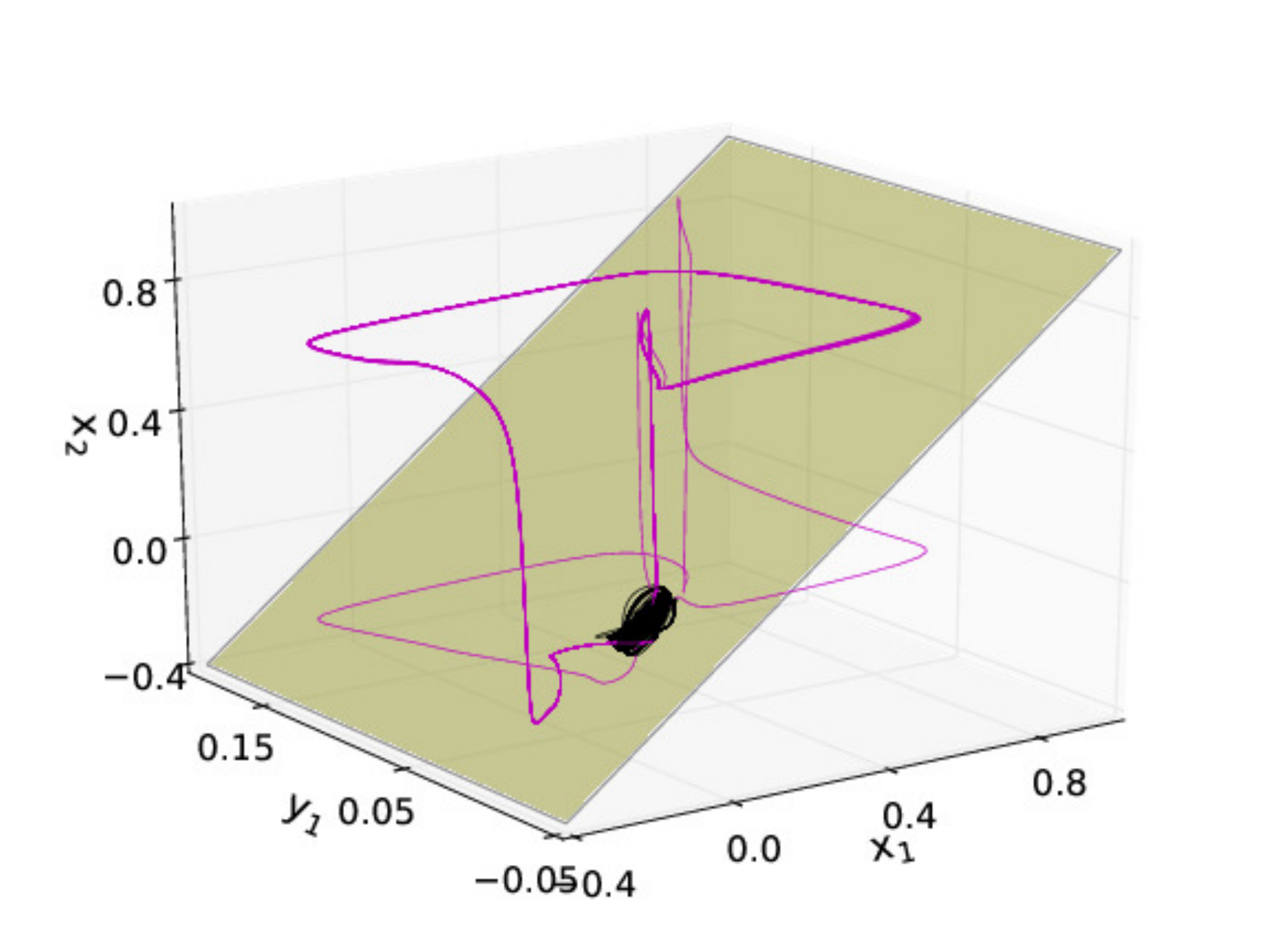}
\caption{(Color Online) Various representations of the long-term dynamics of two FHN units coupled with a single delay when the synchronization manifold is unstable: the top-left panel: time-evolution of the $x$-coordinates of the oscillators; top-right panel: trajectories of the two oscillators in phase space; bottom panel trajectory in a 3-dimensional representation. The plane is the synchronization manifold. Color code: Small amplitude oscillations in black; asynchronous large amplitude oscillations in magenta (dark gray). Parameters: $M=0.01$, $\tau=70$.}
\label{fig:SingleSmallDelay}
\end{figure}

\begin{figure}
\includegraphics[width=\linewidth]{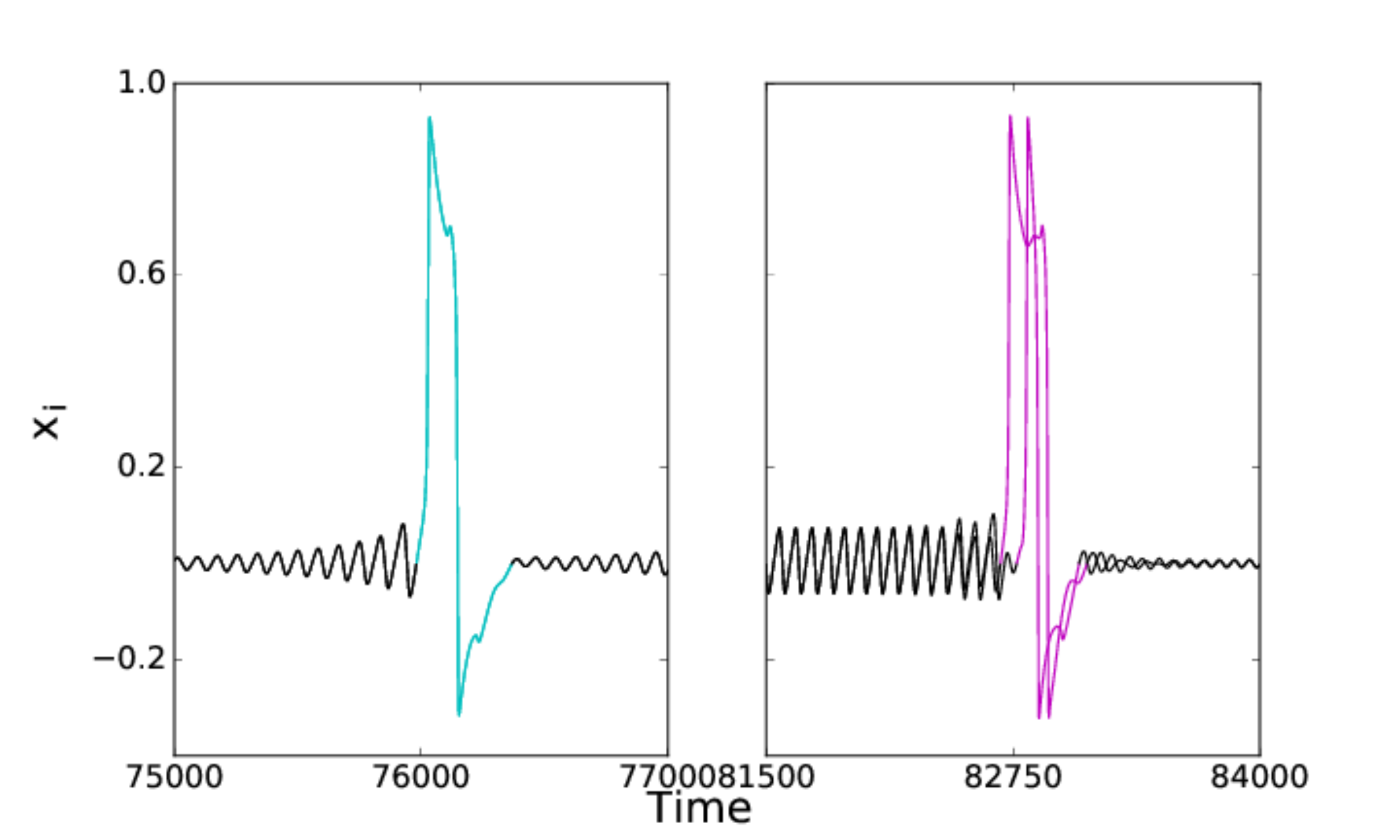}
\caption{(Color Online) Close-up view of typical time series showing in-phase (left panel) and out-of-phase (right panel) events. Color code: Small amplitude oscillations in black; synchronous large amplitude oscillations in cyan (light gray); asynchronous large amplitude oscillations in magenta (dark gray). Parameters: $M=0.01$, $\tau=80$ for the left panel; $M=0.01$, $\tau=70$ for the right panel.}
\label{fig:TwoTypes}
\end{figure}

If the delay is reduced to $\tau=70$ and the coupling strength kept at $M=0.01$, the synchronization manifold is found to be transversally unstable. Therefore, trajectories starting away from the synchronization manifold no longer converge to it. Nonetheless, they spend long time intervals near that manifold: A typical trajectory which does not start on the invariant synchronization manifold is attracted towards it. When it comes sufficiently close to the manifold, it starts executing small amplitude oscillations (see Fig.~\ref{fig:SingleSmallDelay}). After some oscillations, the system is ejected far away from the invariant manifold and it executes a large amplitude oscillation away from the synchronization manifold. At the end of this oscillation, the trajectory returns to the neighborhood of the manifold near the origin again executing small amplitude oscillations, spiraling out from the neighborhood of the origin. Since there are no stable fixed points in the system for the chosen parameters, this motion continues forever. Though this dynamics is chaotic, the time spans between two subsequent large amplitude oscillations varies only a little so that the whole dynamics appears as almost periodic with respect to the large excursions in phase space.

Note that although the trajectory described above comprises both, small and large amplitude oscillations, it is qualitatively different from the typical trajectory described in the previous subsection where the synchronization manifold is stable. When the synchronization manifold is transversally stable, the entire long-term dynamics of the system occurs on that manifold. By contrast, when the synchronization manifold is transversally unstable, the trajectory never reaches that manifold. This implies that even during the small amplitude oscillations, when the system is very close to synchrony, there is a finite separation between the trajectory and the synchronization manifold. More importantly, this difference does not monotonically decrease if the system is allowed to evolve for extremely long time spans. The difference between the two cases is even more evident when comparing the large amplitude oscillations. Since the large amplitude oscillations are on the invariant synchronization manifold when it is transversally stable, the two oscillators are in complete synchrony during the oscillation. By contrast, when the manifold is transversally unstable, the oscillators loose their nearly synchronized state during the large amplitude oscillation entirely. The excursion is so far away from the synchronization manifold, that it is clearly seen in the time series plots as two different shapes for the two oscillators (see Fig.~\ref{fig:TwoTypes}). The oscillation in one of the units starts before and ends after the oscillation in the other unit. The duration of the large amplitude oscillation in the two units is therefore different.

Finally we note that trajectories starting precisely on the synchronization manifold converge to a low-amplitude limit cycle residing on that manifold. Since every trajectory starting on the manifold converges to the limit cycle and no trajectory starting away from the manifold converges to it, we conclude that this limit cycle is stable along the synchronization manifold and unstable transverse to it. This transverse instability of the limit cycle causes the ejection of trajectories that do not start on the synchronization manifold.

If the coupling strength is increased, such that there is additionally a pair of stable fixed points in the system, any trajectory which starts away from the invariant manifold converges finally to one of the two fixed points. However, the general characteristics of the very long transients observed before convergence to the fixed points is similar to the one described above in this subsection.

\subsection{Two Delays}

\begin{figure}
  \includegraphics[width=0.9\linewidth]{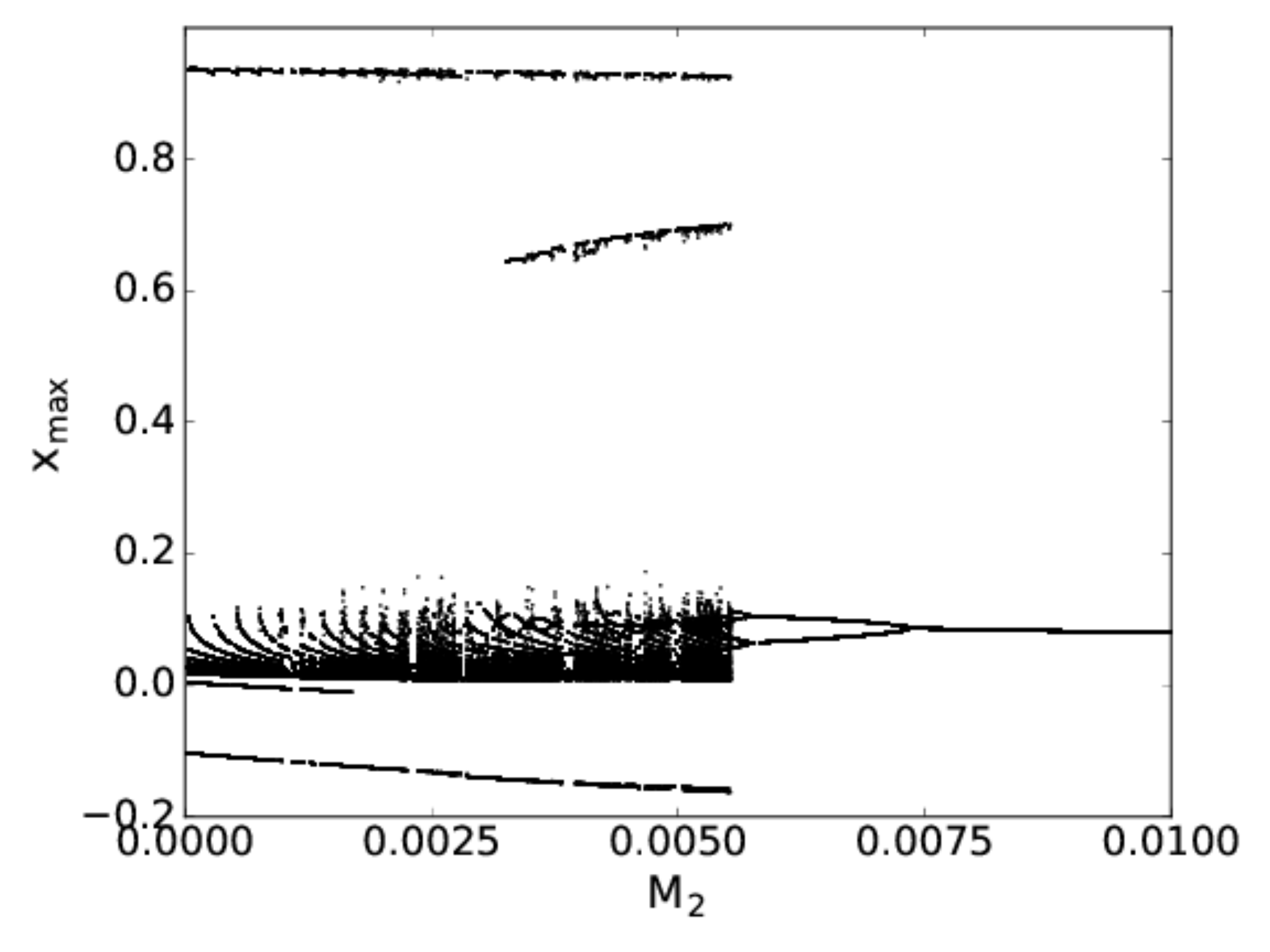}
  \includegraphics[width=0.9\linewidth]{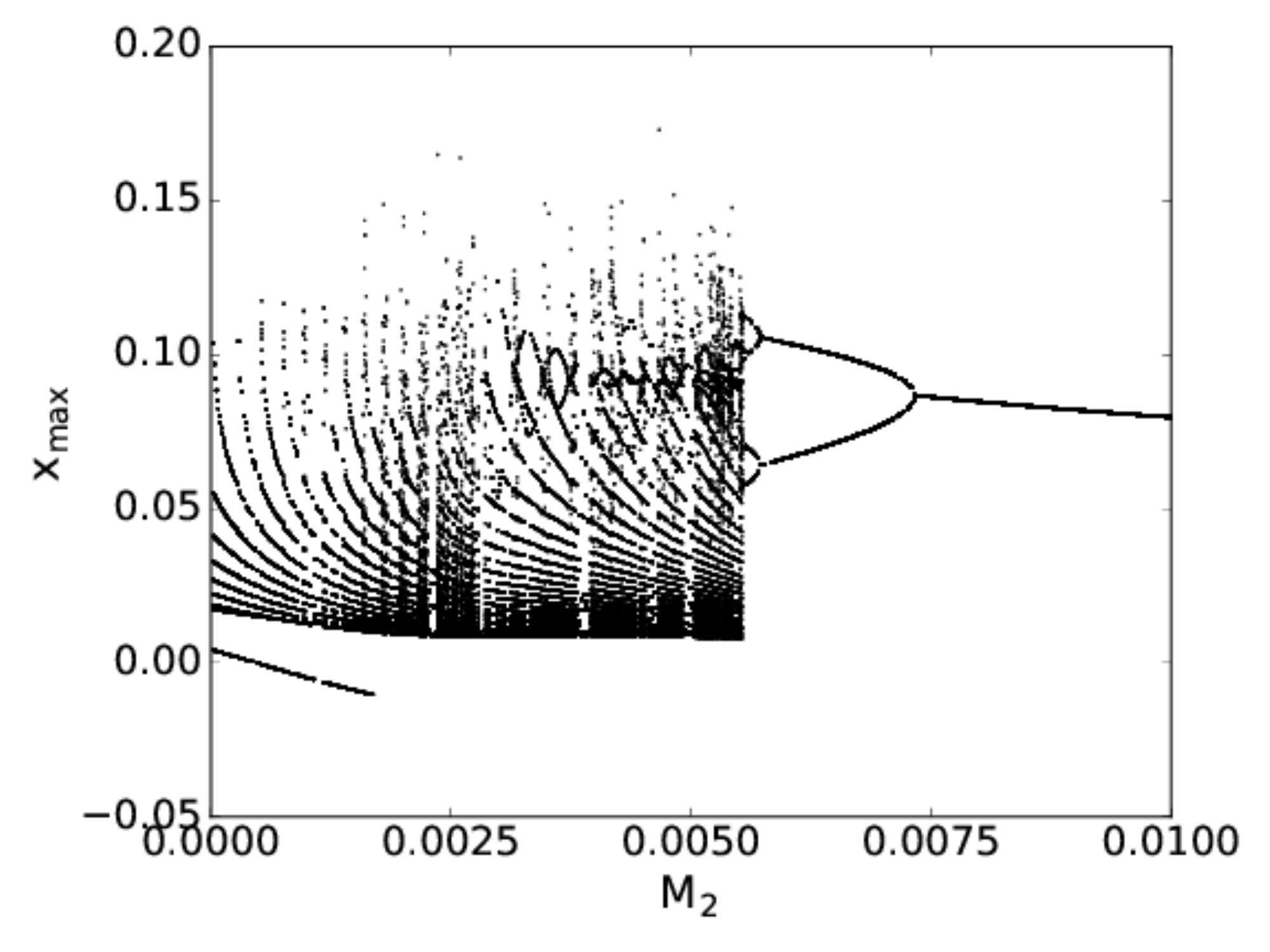}
  \caption{Bifurcation diagram outlining the changes in the structure of the invariant sets on the synchronization manifold with varying coupling strength $M_2$. The vertical axis corresponds to the maximas of the oscillations. Upper panel: complete diagram. Lower panel: close-up view of the region corresponding to small amplitude oscillations. Fixed parameters: $\tau_1=80$, $M_1=0.005$, $\tau=70$}
\label{fig:Bifurcation}
\end{figure}

\begin{figure}
  \includegraphics[width=\linewidth]{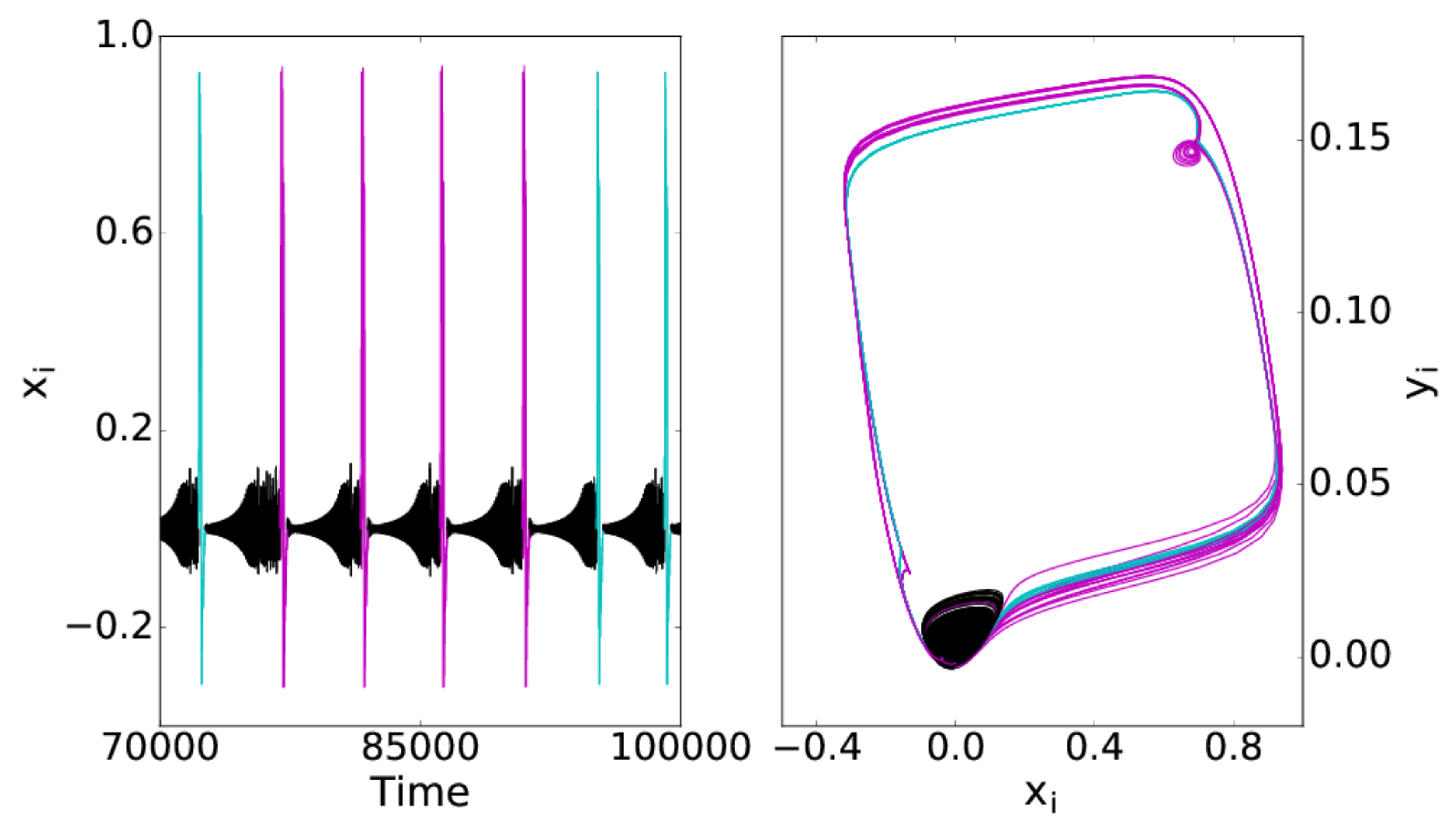}
  \includegraphics[width=\linewidth]{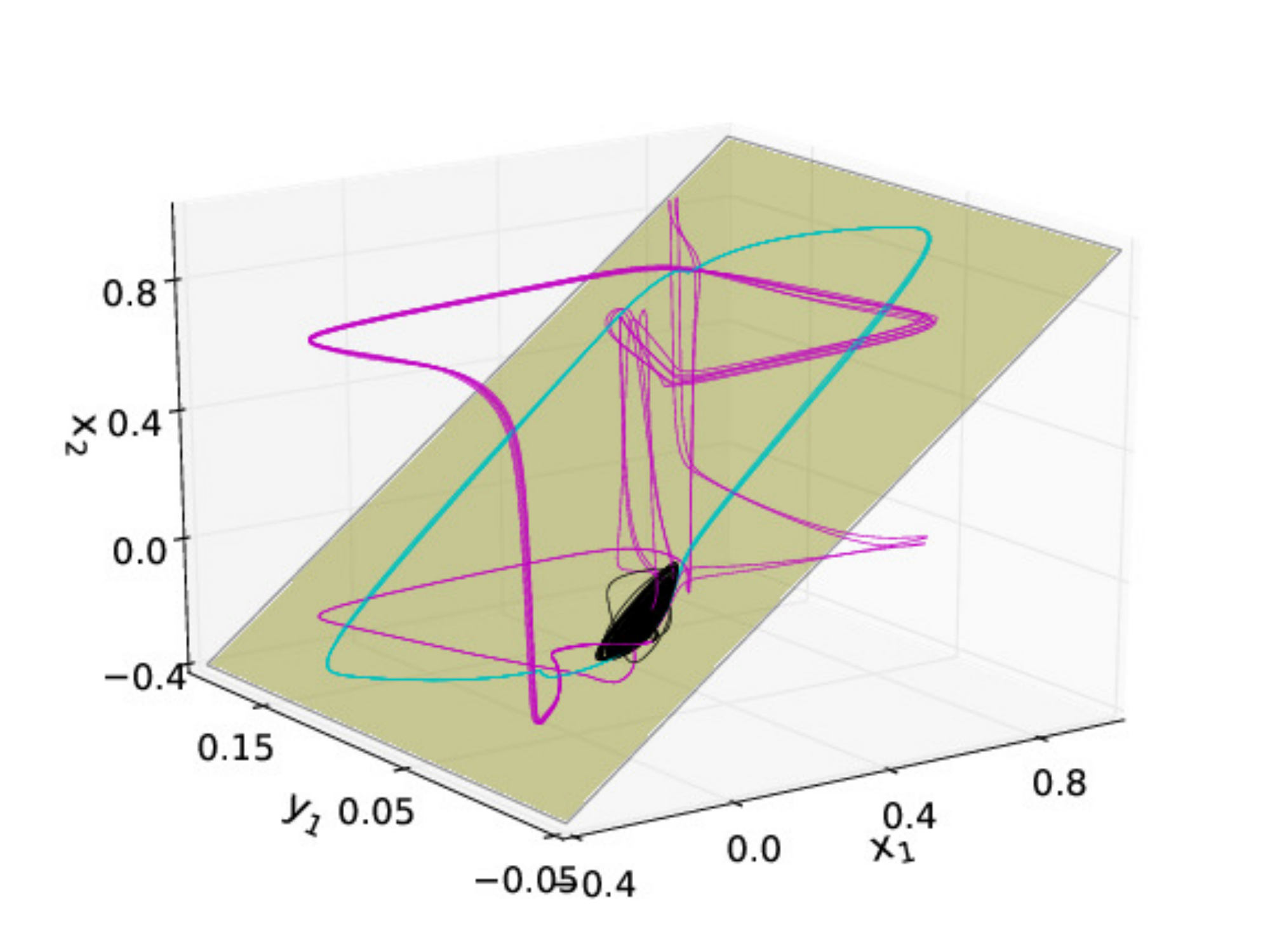}
  \caption{(Color Online) Various representations of the long-term dynamics of two FHN units coupled with two delays: top-left panel: time-evolution of the $x$-coordinates of the oscillators; top-right panel: trajectories of the two oscillators in the phase space; bottom panel: trajectory in a 3-dimensional representation. The plane is the synchronization manifold. Color code: Small amplitude oscillations in black; synchronous large amplitude oscillations in cyan (light gray); asynchronous large amplitude oscillations in magenta (dark gray). Parameters: $M_1=0.005$, $\tau_1=80$; $M_2=0.0053$, $\tau_2=70$.}
\label{fig:TwoDelays1}
\end{figure}

Having analyzed the dynamics of FHN units coupled using a single delay, we now add another coupling with a different delay to the system and compare our results with the previous case. Writing Eq.~\eqref{eq:Model} for $L=2$, we have,
\begin{align}
\dot{x}_i &= x_i (x_i-1) (a-x_i) - y_i + M_1 \left(x_j^{(\tau_1)}-x_i \right) \nonumber \\
&+ M_2 \left(x_j^{(\tau_2)}-x_i\right)\\
\dot{y}_i &= bx_i - cy_i +  M_1 \left(y_j^{(\tau_1)}-y_i\right) + M_2 \left(y_j^{(\tau_2)}-y_i\right)
\end{align}
where $i,j \in \left\{ 1,2 \right\}$. This system can exhibit extreme events if the coupling parameters $M_1$, $M_2$, $\tau_1$ and $\tau_2$ are chosen appropriately. For such a choice of parameters, the dynamics exhibits a combination of properties similar to both the single delay cases considered earlier.

In order to analyze the system, we fix three of the four parameters at $\tau_1=80$, $M_1=0.005$ and $\tau_2=70$. We then vary $M_2$ and note the qualitative changes in the dynamics of the system (see Fig.~\ref{fig:Bifurcation}). If $M_2=0$, the only effective delay in the system is $\tau_1=80$ with total coupling strength $M=M_1=0.005$. Hence the oscillators execute synchronized mixed mode oscillations on the synchronization manifold identical to the single delay case. As $M_2$ is increased to small non-zero values, i.e.\ computing the bifurcation diagram in Fig.~\ref{fig:Bifurcation} starting from its left end, the dynamics differs quantitatively from that of the single delay case. However, the qualitative properties remain similar to the single delay system with a transversally stable synchronization manifold --- that is, the system undergoes a period adding cascade interspersed with chaotic windows up to $M_2 \approx 0.0048$.

By contrast, when we start computing the bifurcation diagram, Fig.~\ref{fig:Bifurcation}, from its right end at large values of $M_2$ and subsequently decreasing $M_2$, we observe a qualitative behavior of the system similar to the single delay system at $\tau=70$, exhibiting a transversally unstable synchronization manifold. As in the single delay case, the transverse instability of the synchronization manifold is due to the presence of a transversally unstable limit cycle located on the synchronization manifold. Therefore, trajectories starting away from the manifold execute small in-phase and large out-of-phase oscillations, while trajectories starting on the synchronization manifold get attracted to this limit cycle as it is stable along the manifold. As $M_2$ is gradually decreased, this limit cycle undergoes a series of period doubling bifurcations.

As $M_2$ is decreased beyond $M_2 \approx 0.0058$, two distinct changes occur in the structure of phase space: The synchronization manifold gains transverse stability; and the transversally unstable limit cycle looses its stability along the synchronization manifold. Due to the loss of stability along the synchronization manifold, the transversally unstable limit cycle becomes inaccessible for trajectories starting on the synchronization manifold. These trajectories now converge to the stable chaotic attractor on the synchronization manifold formed by the period adding cascade. The unstable limit cycles formed due to period doubling bifurcations get embedded in the stable chaotic attractor. This leads to the creation of an intricate dynamics for trajectories starting away from the manifold which is illustrated in Fig.~\ref{fig:TwoDelays1} and described as follows: First, the trajectories are attracted towards the synchronization manifold and execute small amplitude oscillations near that manifold imitating the dynamics of the chaotic attractor. During these oscillations, if the trajectory approaches the vicinity of the embedded transversally unstable limit cycle, it is ejected away from the synchronization manifold, and it executes an out-of-phase large amplitude oscillation; leading to an `out-of-phase' event. The other possibility is that the trajectory avoids the vicinity of the transversally unstable limit cycle sufficiently long during the chaotic small amplitude oscillations and executes a large amplitude oscillation while still being close to the stable chaotic attractor on the synchronization manifold. This case leads to the formation of an `in-phase' event. This dynamical behavior continues until the trajectory has converged to the stable chaotic attractor on the synchronization manifold and finally only `in-phase' events are observed. This rich dynamics can be seen up to $M_2 \approx 0.0048$ beyond which the chaotic attractor ceases to exist and stable mixed mode oscillations are observed.

\begin{figure}
  \includegraphics[width=0.9\linewidth]{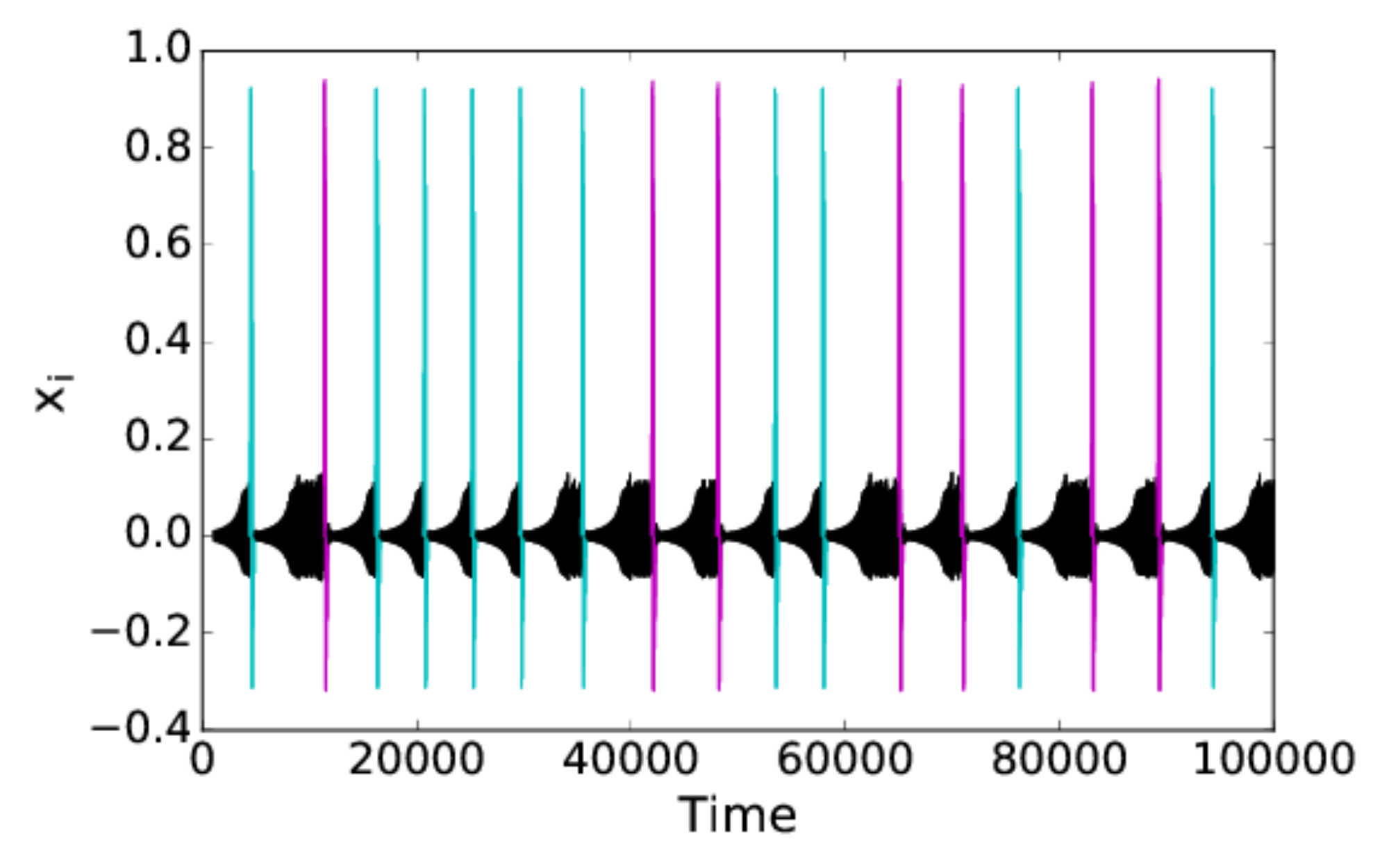}
  \includegraphics[width=0.9\linewidth]{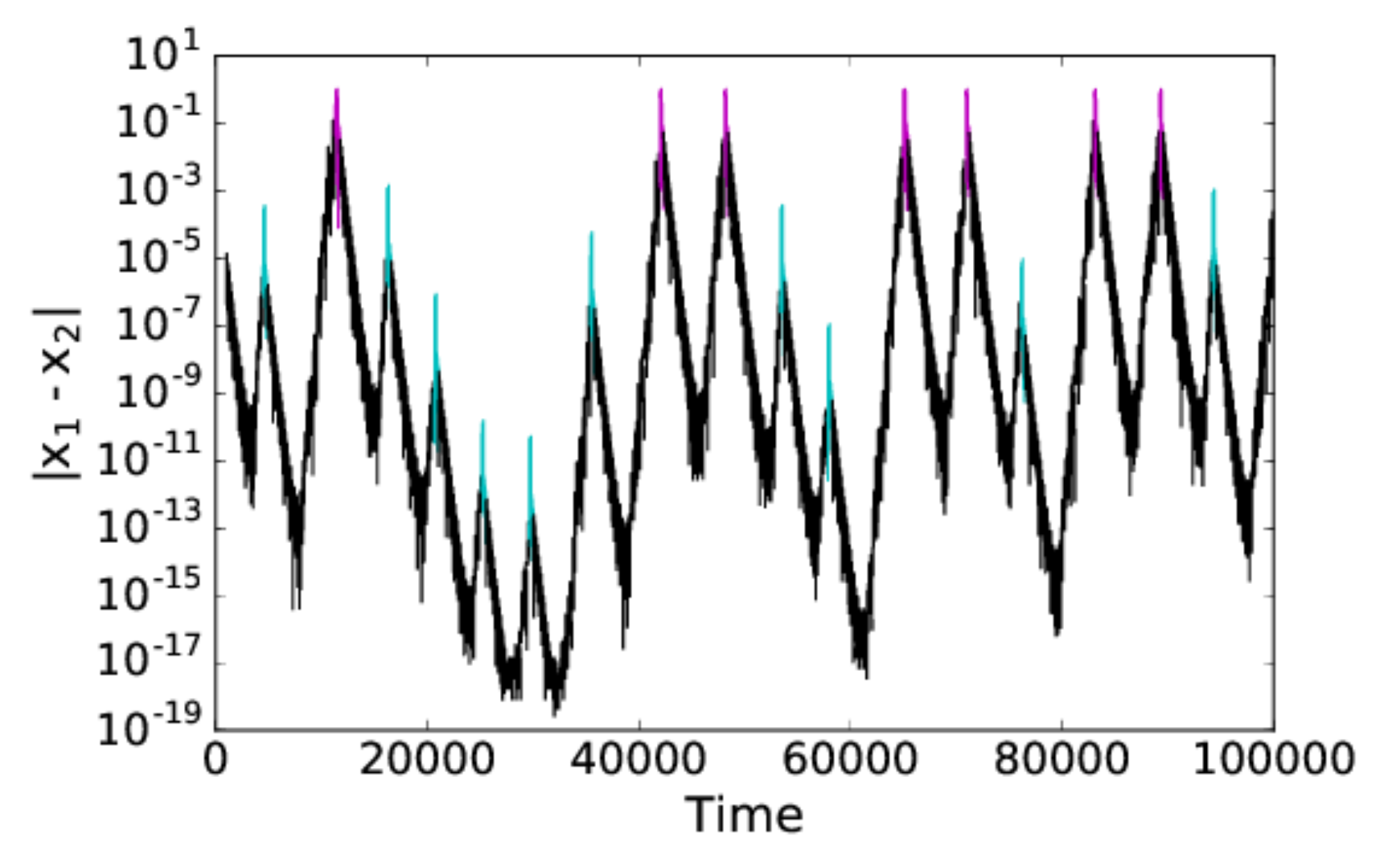}
  \includegraphics[width=0.9\linewidth]{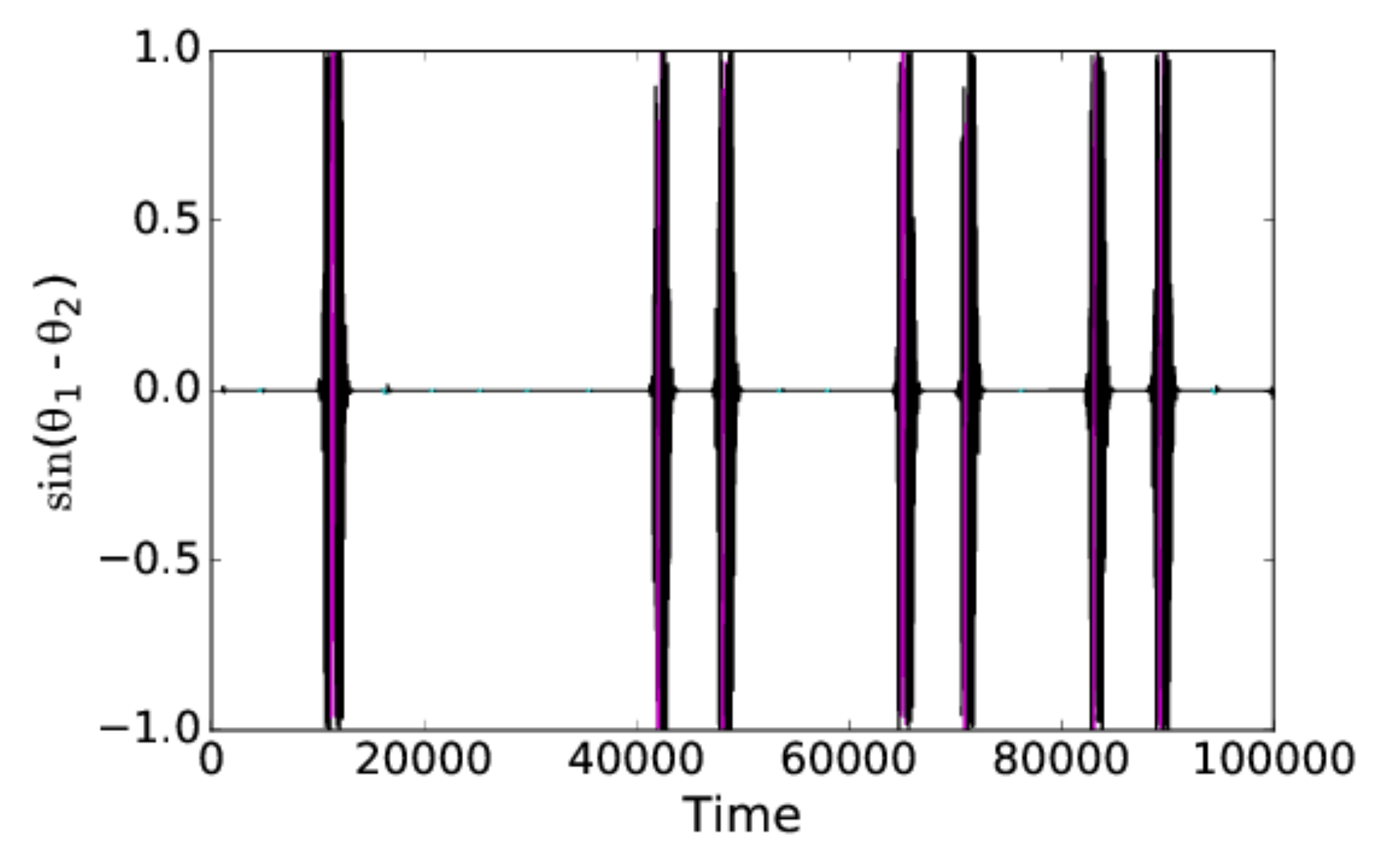}
  \caption{(Color Online) Loss of phase synchrony much before each `out-of-phase' extreme event: top panel: typical time series; middle panel: repeated approach to the synchronization manifold; bottom panel: sine of differences in phases of the oscillators. Color code: Small amplitude oscillations in black; synchronous large amplitude oscillations in cyan (light gray); asynchronous large amplitude oscillations in magenta (dark gray). Parameters: $M_1=0.005$, $M_2=0.0053$, $\tau_1=80$ and $\tau_2=70$.}
\label{fig:Precursors}
\end{figure}

To further illustrate the irregular switching between in-phase and out-of-phase events we present in Fig.\ref{fig:Precursors}, a longer time series together with the corresponding time evolution of the distance of excitatory variable from the synchronization manifold $|x_1-x_2|$ and the differences in the phases of the oscillators $\sin \left( \theta_1-\theta_2 \right)$ with $\theta_i = \arctan \left( \frac{y_i}{x_i} \right)$. We note that a large deviation from the synchronization manifold \textbf{S} happens only during out-of-phase events while in-phase events stay close to \textbf{S}. Moreover, whenever an out-of-phase event occurs, the phase synchrony is lost well before that event and is restored only a long time after it.

\begin{figure*}
  \centering
  \includegraphics[width=\textwidth]{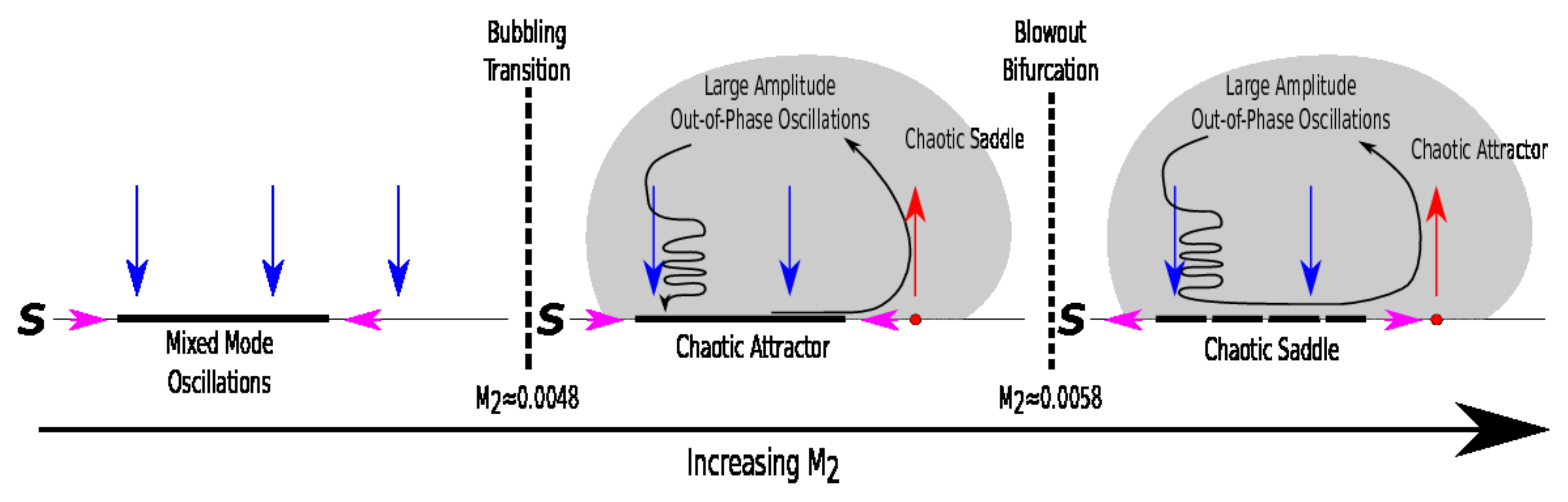}
  \caption{(Color Online) Sketch representing the mechanisms of transition between the three major dynamical regimes obtained by varying $M_2$. For all three panels in the diagram, the synchronization manifold \textbf{S} of the system is represented by the dotted horizontal line. The transversally unstable limit cycle responsible for the ejection of trajectories away from the synchronization manifold is represented by a solid red (dark gray) dot on the synchronization manifold. All other non-trivial invariant subsets on the manifold \textbf{S} are represented by a thick black line which is solid (dashed) if the invariant subset is stable (unstable). Note that the periodic orbits which are transversally unstable are in fact embedded in the chaotic attractor but shown outside it for better visibility. Chaotic attractors and chaotic saddles which are not on the synchronization manifold are represented by the gray region. Attracting and repelling directions are represented by straight colored arrows and the curved or wiggly black arrows represent the trajectories in phase space.}
\label{fig:Mechanism}
\end{figure*}

The observed dynamics in this parameter range can be explained by the appearance of a bubbling transition~\cite{venkataramani1996transitions, venkataramani1996bubbling, ashwin1996attractor, ashwin1994bubbling} and a blowout bifurcation~\cite{ott1994blowout, ashwin1998unfolding}. A sketch of the bifurcation sequence is presented in Fig.~\ref{fig:Mechanism}. For small $M_2$ the synchronization manifold is transversally stable and the only attractor is a mixed mode oscillation located in that manifold (see Fig.~\ref{fig:Mechanism} left panel). With increasing $M_2$ we observe the appearance of a chaotic attractor within the synchronization manifold (see Fig.~\ref{fig:Mechanism} middle panel). At $M_2 \approx 0.0048$ a bubbling transition occurs leading to a loss of transverse stability of one (and subsequently more and more) unstable periodic orbit(s) in the synchronization manifold. The invariant set within the manifold is a chaotic attractor, but there exists a measure zero set in the synchronization manifold, starting from which the trajectories would be ejected from it. This loss of transverse stability is connected with the emergence of trajectories containing large out-of-phase excursions. These excursions comprise of small amplitude and large amplitude oscillations which together make up a chaotic saddle outside the synchronization manifold. From the chaotic saddle the trajectories escape along its unstable directions and approach the attractor in the synchronization manifold. Finally they converge to the chaotic attractor (see Fig~\ref{fig:Mechanism} middle panel). The dynamics resulting from this bubbling transition can be characterized as an in-out intermittency because the mechanisms of ejection from and approach to the synchronization manifold are different (see Fig.~\ref{fig:Precursors})~\cite{blackbeard2014synchronisation}. A further increase of $M_2$ leads finally to a blowout bifurcation in which the synchronization manifold loses its transverse stability completely and the chaotic saddle outside the synchronization manifold becomes an attractor (see Fig~\ref{fig:Mechanism} right panel). Beyond the blowout bifurcation, we obtain the previously described behavior of chaotic trajectories containing out-of-phase large amplitude oscillations separated by small amplitude oscillations.

\begin{figure}
\includegraphics[width=\linewidth]{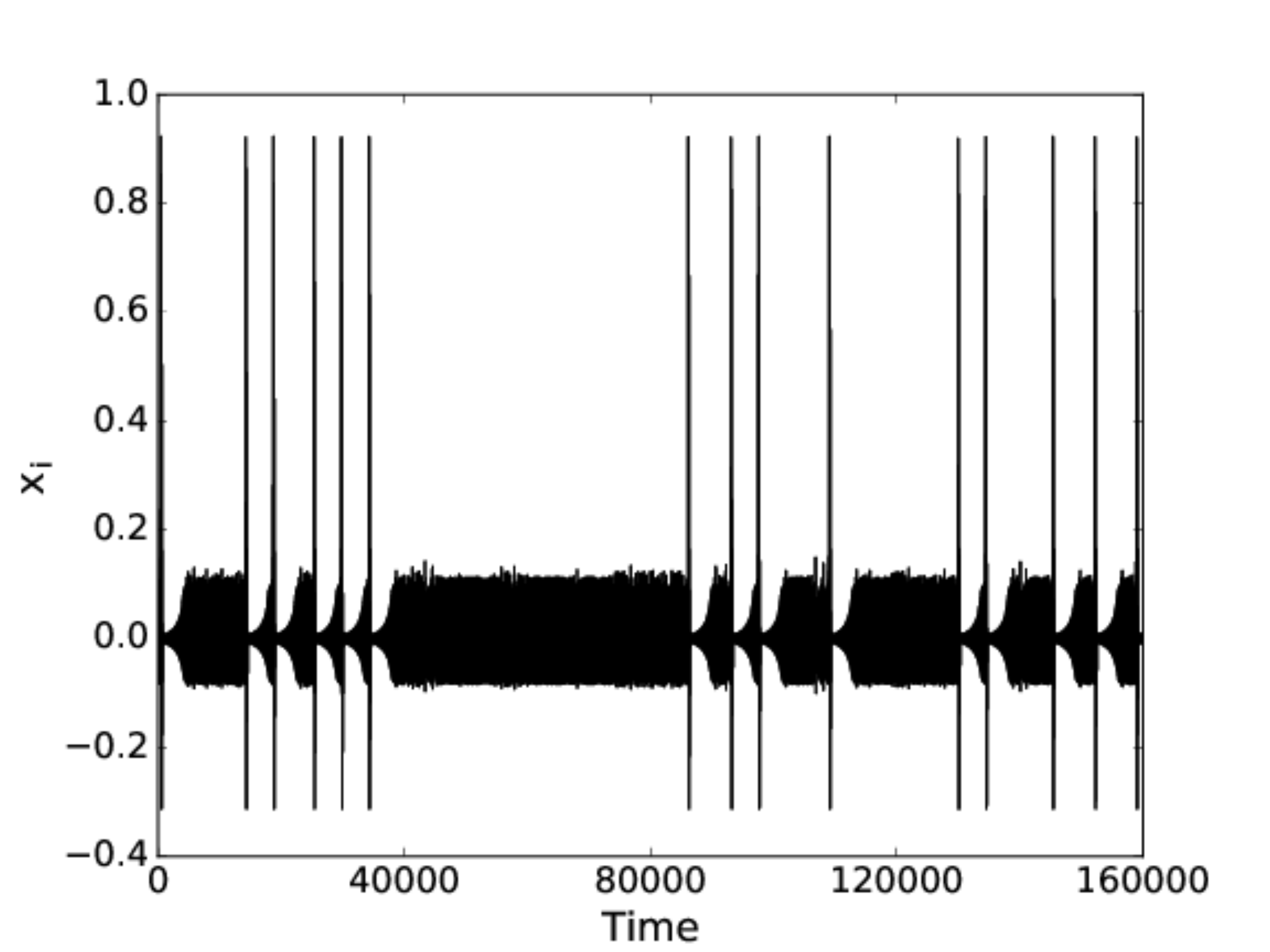}
\includegraphics[width=\linewidth]{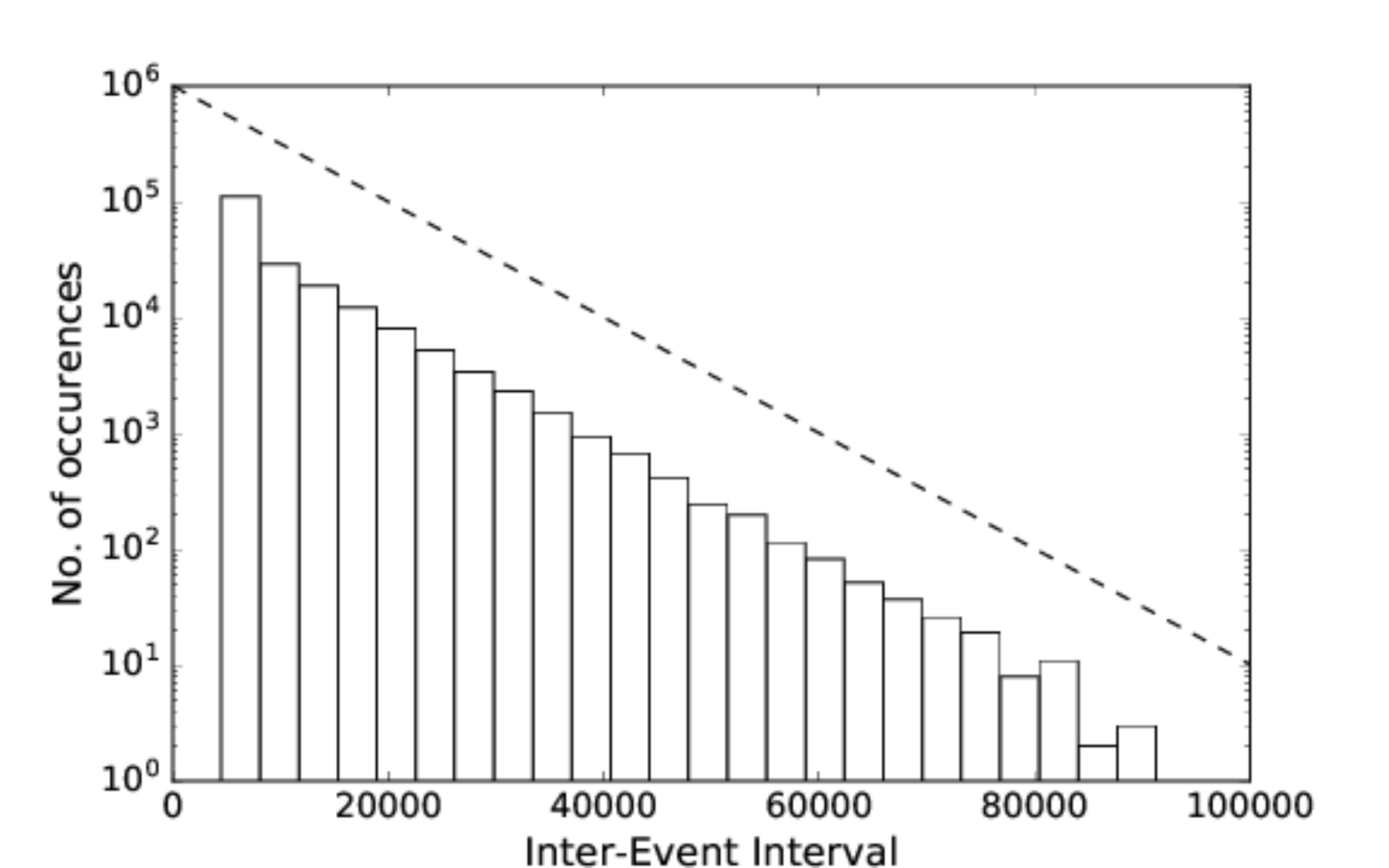}
\caption{Top panel shows part of a typical time series of two FHN units coupled using two delays exhibiting extreme events. Bottom panel shows the histogram of the interevent intervals $t_{IEI}$ for the system. The observation time for the entire simulation is $2 \times 10^9$ time units in which a total of $194924$ events were observed. The dashed line is a multiple of $\exp\left( -r t_{IEI} \right)$ with $r=1.15 \times 10^{-4}$. Parameters: $\tau_1=80$, $M_1=0.01$, $\tau_2=65$ and $M_2=0.00255$.}
\label{fig:TwoDelays}
\end{figure}

So far we have illustrated the intricate interplay between the chaotic attractor, chaotic saddle and the limit cycle only with trajectories for which the occurrence of in-phase and out-of-phase events seems to be almost periodic. However, there are also large fractions of the parameter space shown below where these events occur irregularly and are very rare, so that their interevent intervals are distinctively high. A part of such a trajectory is shown in Fig~\ref{fig:TwoDelays} together with a probability distribution of the interevent intervals. The irregularity of the observed events is manifested by the Poisson-like distribution. Therefore we may call such events to be extreme events because of the rarity of their emergence.

\begin{figure}
\includegraphics[width=\linewidth]{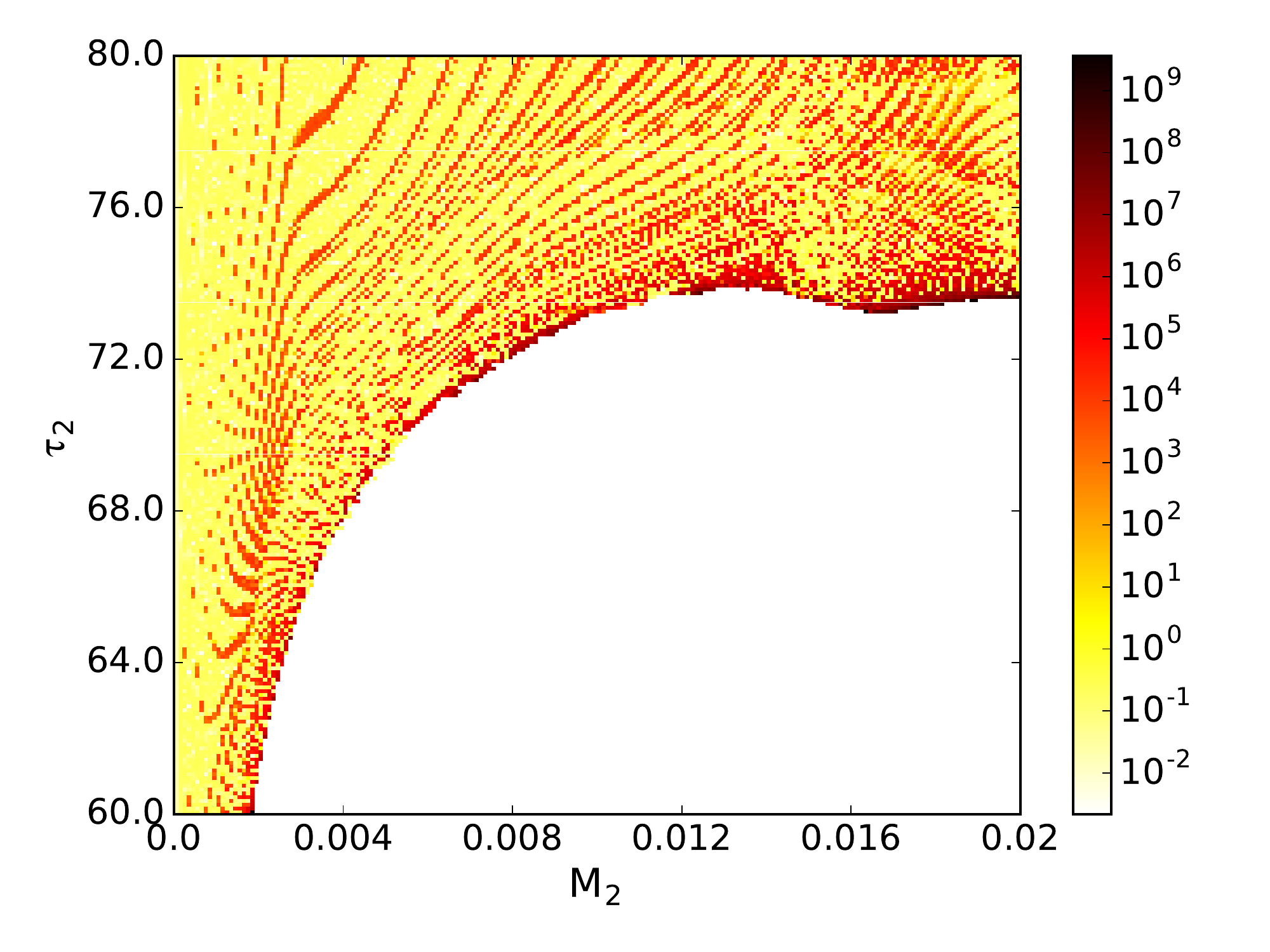}
\caption{(Color Online) Variance of interevent intervals for trajectories starting on the synchronization manifold for varying parameters $\tau_2$ and $M_2$. The white region corresponds to small amplitude oscillations and the brownish (dark gray) region corresponds to the formation of extreme events with very large interevent intervals.}
\label{fig:ISI}
\end{figure}

Until now, we have shown the emergence of extreme events by varying only one of the four coupling parameters of the system. However, extreme events as discussed above occur over a much larger region of parameter space following a very similar mechanism. To illustrate this point, we scan a two-dimensional subsection of the parameter space using the parameters $M_2$ and $\tau_2$ and plot the variance of the interevent intervals (see Fig~\ref{fig:ISI}). Note that in order to highlight the changes in qualitative properties of the invariant synchronization manifold, all trajectories were chosen to start on that manifold. Each pair of parameters $(M_2,\tau_2)$ is colored as follows: If the long-term state of a trajectory starting from a random initial condition on the invariant manifold is such that it exhibits large amplitude oscillations (or events), we compute the variance of the interevent intervals and color the point accordingly. If, on the other hand, the long-term state is such that the entire motion is composed of small amplitude oscillations, then such a variance cannot be computed and the corresponding point is colored white. Note that the regions corresponding to high coupling strength $M_2$ and small time-delays $\tau_2$ exhibit no large amplitude oscillations as they have converged to the transversally unstable limit-cycle on the invariant manifold which is stable parallel to the synchronization manifold. If we now decrease $M_2$ or increase $\tau_2$, the limit-cycle undergoes a period-doubling cascade and subsequently looses its stability parallel to the synchronization manifold. This marks the onset of extreme events, as the trajectories now converge to the chaotic attractor on the invariant manifold and exhibit large amplitude oscillations with high irregularity (as marked by the colored band surrounding the white region). Moving further to the left or the top, we enter the region of the reverse period-adding cascade where we have regions of mixed-mode oscillations interspersed by small chaotic windows similar to Fig~\ref{fig:PeriodAdding}.

\section{Conclusions}

In this study, we have shown that time-delayed coupling with two different delays can cause extreme events in excitable systems by analyzing the dynamical properties of delay-coupled identical FHN oscillators. First we analyzed the dynamics when two FHN units are coupled by a single delay coupling. In this case the FHN oscillators can exhibit a combination of small and large amplitude oscillations either in form of mixed mode oscillations or as chaotic oscillations. We also find that the time delay divides the system into two distinct parameter regimes: If the synchronization manifold is transversally stable (corresponding to larger time delays), the entire long-term dynamics of the system is synchronized. On the other hand, if the synchronization manifold is transversally unstable (corresponding to smaller time delays), the oscillators remain nearly synchronized during the small amplitude oscillations  and are ejected completely out of synchrony during the large amplitude oscillation. We also showed that the ejection is due to the presence of a limit-cycle on the invariant synchronization manifold of the system which is attracting along the manifold but repelling in the directions transverse to it. Though oscillations in such single-delay coupled FHN units were too regular to be classified as extreme events, the dynamical characteristics of the system proved to be crucial to understand the scenario where the FHN units were coupled using two distinct time-delays.

The system consisting of two coupled FHN units with two delays behaves similar to their single-delay counterpart when one of the coupling strengths is largely dominant over the other. However, if the two coupling strengths are comparable, the structures close to or on the synchronization manifold corresponding to both the single-delay regimes interact with each other.  This results in the system exhibiting dynamical properties related to both single-delay regimes. In particular, the large amplitude oscillations can irregularly switch between being in- or out-of-phase. It also results in the small amplitude oscillations becoming chaotic. Furthermore we find regions in parameter space spanned by the second time delay and the corresponding coupling strength where interevent intervals, i.e.\ the time intervals between subsequent large amplitude oscillations exhibit a very large variability including very rare events.

Furthermore we identified the mechanism explaining the emergence of these extreme events. The emergence of extreme events occurs where the period-adding cascade of mixed-mode oscillations located on the synchronization manifold meets the period-doubling cascade of a limit cycle which also lies in the synchronization manifold but is transversally unstable. In the parameter region where both cascades meet, we find a strip in which extreme events occur. This strip is bounded on one side by a bubbling transition where a limit cycle in the synchronization manifold loses its transverse stability and a blowout bifurcation where the whole synchronization manifold becomes transversally unstable. Within this parameter range, we find a chaotic saddle comprising small amplitude oscillations as well as in-phase and out-of-phase large amplitude oscillations alternating irregularly. This very long transient chaotic dynamics finally converges to a chaotic attractor located on the synchronization manifold. The alteration between in-phase and out-of-phase dynamics can be interpreted as in-out intermittency as introduced by Blackbeard et al.~\cite{blackbeard2014synchronisation} because the mechanisms of approaching to and ejection from the synchronization manifold are different.

Finally, we would like to point out that our findings are similar to the ones by Flunkert et al.~\cite{flunkert2009bubbling} who showed the occurrence of a bubbling transition in relay-coupled lasers. The relay-coupling used acts effectively like a delay. Since they have used only a single delay for the coupling their dynamics does not include the formation of extreme events and the irregular alteration between in-phase and out-of-phase events. For the latter, two different delays appear to be a necessary condition. Extreme events as a result of bubbling transitions, as demonstrated here, have also been found by Cavalcante et al.~\cite{gauthier2013predictability} in a system of two diffusively coupled electronic circuits. The difference between their and our results lies in the fact that the sizes of the extreme events in the electronic system possess a distribution function characteristic for dragon-kings. By contrast, our time delay coupled FHN model does not exhibit such a distribution because the sizes of the extreme events are almost equal. Experimentally, extreme events containing chaotic small amplitude oscillations and large amplitude events have been found in a diode laser subject to a phase-conjugate feedback~\cite{dal2013extreme} and in a mode-locked fiber ring laser~\cite{lecaplain2012dissipative}. Though the experimental setups do not consider a coupling between two laser systems as studied here, they show some similarities to the dynamics studied.

\section*{Acknowledgments}
The authors would like to thank G. Ansmann, P. Ashwin, A. Choudhary, P. H\"ovel, E. Knobloch, K. Lehnertz, C. Masoller, E. Sch\"oll and S. Wieczorek for fruitful discussions and critical suggestions. This work was supported by the Volkswagen Foundation (Grants No. 88459). The simulations were performed at the HPC Cluster CARL, located at the University of Oldenburg (Germany) and funded by the DFG through its Major Research Instrumentation Programs (INST 184/157-1 FUGG) and the Ministry of Science and Culture (MWK) of the Lower Saxony State. 

\bibliography{Ref}

\end{document}